\documentclass[fleqn,usenatbib]{mnras}

\usepackage[T1]{fontenc}

\DeclareRobustCommand{\VAN}[3]{#2}
\let\VANthebibliography\thebibliography
\def\thebibliography{\DeclareRobustCommand{\VAN}[3]{##3}\VANthebibliography}

\usepackage{graphicx}	% Including Fig.~files
\usepackage{amsmath}	% Advanced maths commands
\usepackage{amssymb}	% Extra maths symbols

\usepackage{placeins}   % Confine floats to section
\usepackage[caption=false]{subfig} % Subfigures
\usepackage{cuted}      % onecolumn equation
\usepackage{flushend}   % onecolumn equation

\usepackage{newtxtext,newtxmath}
\title[Spin-based removal of instrumental systematics]{Spin-based removal of instrumental systematics in 21cm intensity \\
mapping surveys}

\author[N. McCallum et al.]{
Nialh McCallum,$^{1}$
\thanks{E-mail: nialh.mccallum@postgrad.manchester.ac.uk}
Daniel B. Thomas,$^{2}$
\thanks{E-mail: dan.b.thomas1@gmail.com}
Philip Bull,$^{2,3}$
and Michael L. Brown$^{1}$
\\
$^{1}$Jodrell Bank Centre for Astrophysics, School of Physics \& Astronomy, The University of Manchester, Manchester M13 9PL, UK
\\
$^{2}$School of Physics and Astronomy, Queen Mary University of London, London, E1 4NS, UK
\\
$^{3}$Department of Physics and Astronomy, University of Western Cape, Cape Town 7535, South Africa
\\
\\
}

\date{Accepted XXX. Received YYY; in original form ZZZ}

\pubyear{2021}

\begin{document}
\label{firstpage}
\pagerange{\pageref{firstpage}--\pageref{lastpage}}
\maketitle

\begin{abstract}
Upcoming cosmological intensity mapping surveys will open new windows on the Universe, but they must first overcome a number of significant systematic effects, including polarization leakage. We present a formalism that uses scan strategy information to model the effect of different instrumental systematics on the recovered cosmological intensity signal for `single-dish' (autocorrelation) surveys. This modelling classifies different systematics according to their spin symmetry, making it particularly relevant for dealing with polarization leakage. We show how to use this formalism to calculate the expected contamination from different systematics as a function of the scanning strategy. Most importantly, we show how systematics can be disentangled from the intensity signal based on their spin properties via map-making. We illustrate this, using a set of toy models, for some simple instrumental systematics, demonstrating the ability to significantly reduce the contamination to the observed intensity signal. Crucially, unlike existing foreground removal techniques, this approach works for signals that are non-smooth in frequency, e.g. polarized foregrounds. These map-making approaches are simple to apply and represent an orthogonal and complementary approach to existing techniques for removing systematics from upcoming 21cm intensity mapping surveys.
\end{abstract}

\begin{keywords}
cosmology: observations -- large-scale structure of Universe -- radio lines: general -- methods: observational
\end{keywords}

%%%%%%%%%%%%%%%%%%%%%%%%%%%%%%%%%%%%%%%%%%%%%%%%%%

%%%%%%%%%%%%%%%%% BODY OF PAPER %%%%%%%%%%%%%%%%%%

\section{Introduction}
 Line intensity mapping, particularly of the 21cm line of neutral hydrogen (HI), promises to develop into an exciting new probe of cosmology \citep{2008MNRAS.383..606W,2008MNRAS.383.1195W,2008PhRvL.100i1303C,2008PhRvL.100p1301L}. This method uses spatial fluctuations in the intensity of spectral line emission to trace fluctuations in the underlying cosmic density field (at low redshift and during the cosmic Dark Ages) and the ionization state of the inter-galactic medium (IGM; during Cosmic Dawn and the Epoch of Reionization) \citep{PhysRevD.78.103511,PhysRevD.78.023529,2009astro2010S.234P,2010MNRAS.407..567B,2010ApJ...721..164S}. %\textcolor{red}{The frequency dependence means that they can be used to probe the redshift distribution along the line of sight, offering a 3D probe of the large-scale structure of the Universe.}
 At redshifts where the spectral line intensity acts as a biased tracer of the dark matter distribution, this facilitates the investigation of the large-scale structure through observations of baryon acoustic oscillations and redshift space distortions, while at redshifts where the ionization state of the IGM is evolving, measurements can be made of the optical depth to reionization and the clustering properties of early ionizing radiation sources \citep{2011ApJ...741...70L,2012A&A...540A.129A,2013MNRAS.434.1239B,2015ApJ...803...21B,2019BAAS...51c.101K}.

A number of experiments have made important advances towards a positive detection of the 21cm line intensity signal, with the most significant progress to date coming from detections in cross-correlation with optical data \citep{2010Natur.466..463C,2013ApJ...763L..20M,2017MNRAS.464.4938W,2018MNRAS.476.3382A,2021arXiv210204946W}. There has not yet been a definitive detection of the HI auto-power spectrum however, although future experiments such as the SKA \citep{2015aska.confE..19S}, along with pathfinders such as MeerKAT \citep{2017arXiv170906099S}, have the potential to make a significant first detection. %With the upcoming dedicated surveys and its sensitivity to a variety of phenomena, intensity mapping promises to be an important cosmological probe in the coming era.

There are a number of obstacles to tackle in order to definitively measure the HI signal. Foremost is the fact that the 21cm signal is far fainter than many of the foreground sources. Fortunately the unpolarized foreground frequency spectra tend to be relatively smooth, and as such component separation techniques should be capable of isolating the part of the true cosmological signal that resides on smaller radial scales (higher $k_\parallel$ modes) \citep{2015MNRAS.447..400A,2020arXiv201002907C,2020MNRAS.499..304C}. Foreground removal is further complicated by the control of systematics e.g. by calibration -- even if the smooth components are successfully removed to a high level of accuracy, foreground emission could still leak into the final data due to modulation and mixing caused by systematic effects that are non-smooth in frequency. In addition, a number of the foreground sources are polarized -- for example the Galactic synchrotron emission -- and do not have an entirely smooth dependence on frequency due to Faraday rotation and other line-of-sight mixing effects \citep{1986rpa..book.....R,2009A&A...495..697W,2010MNRAS.409.1647J,2013ApJ...769..154M,2014PhRvD..89l3002D}. Despite the polarized foregrounds being a weaker signal than the total intensity foregrounds, they are still much larger than the HI cosmological signal, and as such systematics causing polarization to intensity leakage are a concerning issue for upcoming surveys \cite[e.g.][]{2014A&A...568A.101J,2015MNRAS.451.3709A,2016MNRAS.462.4482A,2018MNRAS.476.3051A,2020arXiv201002907C}.

Future experiments will require careful control of systematics -- especially ones that impart complex spectral structure, such as these -- to reach their desired sensitivities \citep{1996A&AS..117..149S, 2000A&AS..143..515H,2011A&A...527A.106S,2011A&A...527A.107S}. Note that there is also an extensive literature on a variety of other survey systematics, including beam model uncertainties \cite[e.g.][]{2021MNRAS.502.2970A,2021MNRAS.506.5075M}, correlated ($1/f$) receiver noise \cite[e.g.][]{2018MNRAS.478.2416H,2021MNRAS.501.4344L}, and low-level radio-frequency interference (RFI) \cite[e.g.][]{2018MNRAS.479.2024H}. In this paper, we will focus on a subset of instrumental systematic effects that couple the sky signal to the telescope's scanning pattern. Our analysis applies specifically to scanning autocorrelation experiments such as GBT, Parkes, MeerKAT, and SKAO-MID, but not to interferometers. Autocorrelation experiments typically involve steerable individual or multiple dishes that actively scan across the sky, performing direct observations.
%which may involve steerable individual or multiple dishes that actively scan across the sky. Our formalism is not intended to apply to ``drift-scan'' telescopes, typically interferometers, which have a single fixed pointing that maps out the sky over time due to the rotation of the Earth.

In recent years, the Cosmic Microwave Background (CMB) has been the primary cosmological probe and has provided the most stringent measurements of the $\Lambda$-CDM Cosmological model. Consequently it has a rich literature which provides a useful resource for techniques that could be exploited by future intensity mapping surveys. In particular we shall consider how the map-making processes implemented by CMB experiments \citep[e.g.][]{Keihanen:2009tj,10.1111/j.1365-2966.2008.14195.x,2010MNRAS.407.1387S}, whose focus is on polarization and separating it from the intensity, could be extended to aid in the intensity mapping case. It has been pointed out that these map-making methods could also be extended to remove a number of systematic signals \citep[e.g.][]{2014ApJ...794..171P,2015MNRAS.453.2058W,2021MNRAS.501..802M}.

\begin{figure}
  \centering
    \includegraphics[width=0.75\columnwidth]{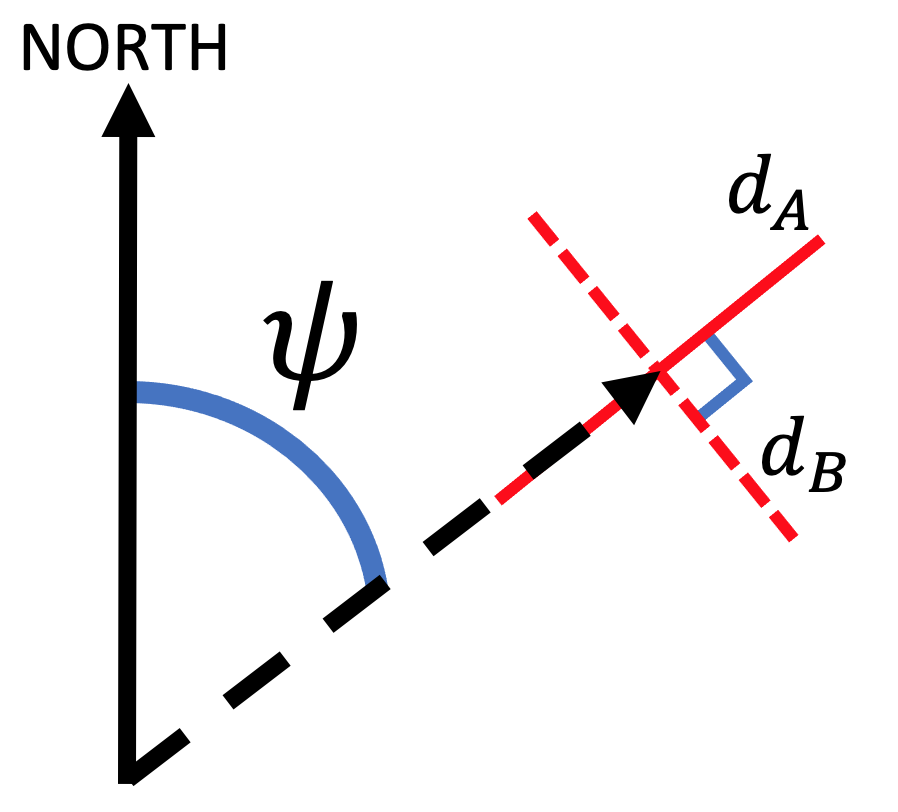}
\caption{Schematic showing the orientation of the linearly polarized feeds $d_A$ and $d_B$ of the instrument with respect to the sky. The crossing angle $\psi$ gives the orientation of the scan direction of the instrument with respect to North.}
\label{figure:CrossingAngle}
\end{figure}

With their focus on polarization, CMB surveys have the inverse problem to intensity mapping -- current experiments are concerned with leakage of the intensity signal into polarization, which is the main target of the current generation of ground-based CMB experiments. Many different techniques and approaches have been used to forecast and address this leakage \citep[e.g.][]{PhysRevD.77.083003,2007MNRAS.376.1767O,2017A&A...598A..25H}, some of which might reasonably also be used to characterize the related effect in intensity mapping surveys. In this work, we will attempt to adapt one such approach to single-dish intensity mapping experiments. In particular, we will apply the methods of \cite{Wallisetal2016}, \cite{2021MNRAS.501..802M,2021arXiv210905038M}, and \cite{2021arXiv210202284T}, which are designed to characterize instrumental systematics and how they couple to the scan strategy based on their {\it spin} properties. Here, spin refers to the dependence of the signals (either ``on-sky'' signals or spurious ones arising from imperfect instrumentation) on the orientation of the instrument with respect to the sky as it scans across a pixel, which we refer to as the crossing angle, $\psi$ as is shown in Fig.~\ref{figure:CrossingAngle}. As we shall see, various systematic effects couple to the scanning strategy in different and particular ways, producing contamination that has different spin structure. By decomposing the observed data into different spin components, we can effectively separate certain instrumental systematic effects from the target cosmological signal more cleanly than would be possible using their angular and spectral structure alone.

In intensity mapping, the targeted cosmological signal is a spin-zero quantity. The spin-zero component of the data can be calculated simply by averaging the data points within a pixel, $_0 S^d=\frac{1}{N}\sum^N_j S(\psi_j,\theta,\phi)$ where $N$ is the number of hits (unique observations) in a pixel, and $S(\psi,\theta,\phi)$ is the measured signal. However, in the presence of instrumental systematics, e.g. dishes that are not ideal, there will be leakage that contaminates the cosmological signal. If this leakage is classified according to its spin, several advantages arise. A key one is the ability to remove the systematics directly with map-making, using the crossing angle information in an analogous way to that used in some CMB map-making processes \citep[e.g.][]{2010MNRAS.407.1387S,2014ApJ...794..171P,2021MNRAS.501..802M}.

This paper is organized as follows. We introduce the spin-based map-making formalism in the context of intensity mapping in Section~\ref{section:Formalism}, and then focus on two case studies of instrumental systematics important in intensity mapping surveys, including the troublesome polarization leakage. We show how to model the case studies for this approach in the context of intensity mapping (Section~\ref{section: Spin Characterisation of Systematics}), finding that using crossing angle information in map-making can effectively remove these systematics (Section~\ref{section:Map-Making}). We conclude in Section~\ref{section:Conclusion}.

\section{Spin formalism for intensity mapping}
\label{section:Formalism}
In this section we present the spin formalism from \cite{Wallisetal2016} and \cite{2021MNRAS.501..802M}, adapting it to apply to single-dish intensity mapping surveys. For clarity, we will present just enough detail for a self-consistent account of how the formalism applies to intensity mapping, while directing to reader to \cite{2021MNRAS.501..802M} for full details of conventions and more detailed derivations -- including the leakage of arbitrary spin fields sourced by systematics to the intensity signal at both map and angular power spectrum level.

We model the signal seen by a single dish at a single frequency from a single observation (scan) as the sum of two orthogonal linearly polarized feeds, $d_A$ and $d_B$, following (e.g.) Eq.~23 of \cite{2020arXiv201113789W}. In the case of an ideal dish, for any intensity, $I(\theta,\phi)$, and polarization, $Q(\theta,\phi)$ and $U(\theta,\phi)$, fields on the sky, the measured signal at location $\theta,\phi$ with crossing angle $\psi$ is
\begin{equation}
\begin{split}
 &S(\psi,\theta,\phi) = \frac{1}{2} [d_A + d_B]
 \\
 &= \frac{1}{2} \Big[\Big ( I(\theta,\phi) + Q(\theta,\phi)\cos(2\psi) + U(\theta,\phi)\sin(2\psi)\Big )\\
& + \Big ( I(\theta,\phi) + Q(\theta,\phi)\cos(2\psi+\pi) + U(\theta,\phi)\sin(2\psi+\pi)\Big )\Big]
 \\
 &= I(\theta,\phi).
 \label{eq:summedsignal}
\end{split}
\end{equation}
As expected, the measured signal for an ideal dish has no $\psi$ dependence -- the linearly polarized feeds are perfectly orthogonal regardless of the direction of the scan, and there is therefore no polarization leakage. As we will show in Section~\ref{section: Spin Characterisation of Systematics}, once we include non-idealities of the dish, the lack of $\psi$ dependence ceases. As well as leaking polarized emission into the unpolarized channel, this can also cause problems in the absence of a polarized sky; our two later case studies include one example of each. Note that all polarization leakage in these surveys is caused by non-idealities of the instrument, so we expect any source of polarization leakage to be amenable to this approach.

For the more general case of a non-ideal dish, we can write the measured signal as a function of the crossing angle $\psi$ and location on the sky $\theta,\phi$ as
\begin{equation}
S(\psi,\theta,\phi)=\sum_{k\geq0}\frac{1}{2}\left({}_{k}\tilde{S}^*(\theta,\phi)e^{ik\psi} + {}_{k}\tilde{S}(\theta,\phi)e^{-ik\psi}\right)\label{eq:decomp}
\end{equation}
 where the ${}_{k}\tilde{S}$ represent a decomposition of the signal into its different spin components,\footnote{This can equivalently be written as $S^Q_k \cos (k\psi)+S^U_k \sin(k\psi)$ with $ _k\tilde{S}=S^Q_k+iS^U_k$.} where we recall that the true cosmological intensity signal contributes only to $k=0$. The detected spin zero signal will now be a combination of the scan strategy and the different spin signals seen by a single dish,
 \begin{eqnarray}
 &&{}_{0}\tilde{S}^{d}(\theta,\phi) = \sum_{k=-\infty}^{\infty} \tilde{h}_{0-k} (\theta,\phi) {}_{k}\tilde{S}(\theta,\phi);\label{eq:SpinSyst}\\
 &&    \tilde{h}_{n} = \frac{1}{N_{\rm hits}}\sum_{j}\left\{\cos(n\psi_{j}) + i \sin(n\psi_{j})\right\},\nonumber\end{eqnarray}
where the $\tilde{h}_{n}$ functions encode the scan strategy within a pixel, $N_{\rm hits}$ is the number of hits within that pixel, and $\{\psi_{j}\}$ are the set of crossing angles associated with those hits. For an ideal scan these quantities will be zero, while in the worst case they are unity, which corresponds to just one distinct crossing angle in a pixel. We should note here that the signals ${}_{k'}\tilde{S}(\theta,\phi)$ will in general be frequency dependent, while the $\tilde{h}_{0-k'} (\theta,\phi)$ will not as these are a function only of the orientation of the telescope.

We will show specific examples of how two instrumental systematics manifest in this formalism in Section~\ref{section: Spin Characterisation of Systematics}. First, we wish to emphasise some important consequences of the formalism for map-making and the scan strategy. These apply to any instrumental systematic for which one can write the signal seen by an individual measurement as a function of the crossing angle $\psi$ as in Eq.~\ref{eq:decomp}, and therefore compute the contamination of the targeted spin zero signal using Eq.~\ref{eq:SpinSyst}. Note that throughout this work we consider a single dish for simplicity, but that the formalism is readily extendable to a multi-dish experiment operating in ``single-dish'' mode; in this case, the $_0S^d$ and $\tilde{h}_n$ will vary between dishes, and the final observed maps can be obtained by averaging the maps made by each dish.

\subsection{Map-making}
If we know the spin $k$ of the contaminating components of $S(\psi,\theta,\phi)$, then we can use the knowledge of the experiment's scan strategy, $\tilde{h}_n$, to solve for spin-$k$ quantities at the same time as constructing $_0 S^d$, by calculating the quantity $\frac{1}{N}\sum^N_j S(\psi_j,\theta,\phi) e^{ik\psi}$. This is no different in principle to the simple binning map-making used by some CMB experiments \citep[e.g][]{2010MNRAS.407.1387S,1997ApJ...480L..87T}. As standard CMB experiments will wish to solve for spin-0 and spin-2 fields when map-making in order to separate the temperature and polarization signals. Extensions to this in order to map-make for systematics signals of other spin have also been considered \citep{2015MNRAS.453.2058W,2021MNRAS.501..802M}. In the intensity mapping case one will always solve for the spin-0 signal at map-making; the choice of which additional spins to include in map-making depends on which dish non-idealities are contributing to $S(\psi,\theta,\phi)$, and what their spins are.

Furthermore, we note that in order to work well, the map-making must be performed on data over timescales which allow little variation in the signature of the systematic effect. If there is significant evolution of the effect between revisits of a pixel then the leaked signals would lose their pure spin dependence and the extended map-making would fail to remove them. In order to combat this case where systematics drift over the time of a full survey, the time ordered data (TOD) would need to be split into shorter timescales over which the systematic is stable -- extended map-making performed on the split data would then remove the systematics. The resulting observed intensity maps would finally be stacked before applying foreground removal.

We will demonstrate this extended map-making technique explicitly for two case studies in Section~\ref{section:Map-Making}. We will show that this technique can effectively remove systematics that are not removed by standard intensity mapping filtering techniques which do not use the crossing angle information from the scanning strategy.

Due to the discrete number of spins that might reasonably be expected to manifest in the data, it may also be possible to use this approach as a ``blind'' way to diagnose systematics. To do this, one would make maps including a further spin field in addition to zero successively. One could then examine the resulting reconstructed spin fields for any signs of systematic signatures that could be corrupting the cosmological signal. For an intensity mapping survey setup, the non-zero spin signals should in principle not be present (appearing noise-like for a realistic survey), with systematics identifiable from the reconstructed signals having a larger than expected contribution. The potential presence of structure emanating from the scanning strategy due to systematic leakage would also act as a diagnostic aid. We present a first look at this approach in Appendix~\ref{section:Blind Diagnosis}, but leave it to future work to examine this diagnostic approach in detail.

\subsection{Scan strategy}
We can use Eq.~\ref{eq:SpinSyst} to calculate the effect that particular instrumental systematics will have on a survey. This is achieved by specifying appropriate forms for the $\tilde{h}_n$ terms for each choice of scan strategy. This also allows us to calculate the extent to which changing the scan strategy can reduce systematics, without requiring any additional analysis steps \citep{Wallisetal2016}.

There are a number of factors that influence scan strategy design, e.g. control of elevation- and azimuth-dependent contributions to the system temperature, or avoiding orbital and terrestrial sources of RFI. These considerations usually dictate that ground-based CMB and intensity mapping surveys should use constant elevation scans (CESs). CESs restrict the range of crossing angles that can occur in the data, placing a fundamental limit on how well the $\tilde{h}_n$ values can be optimized \citep{2021arXiv210202284T}. It is nevertheless possible to provide an idea of the best choices of field and elevation to suppress systematics, for example the polarization leakage. We discuss how this kind of optimization can be done in Appendix~\ref{section:Scanning Strategies}, but a basic summary is that one should choose a strategy that minimizes the value of $|\tilde{h}_n|^2$, averaged over the pixels in the survey, for all $n\ge 1$. This effectively minimizes the coupling of higher-spin components of the observations into the spin-0 map. The $\tilde{h}_n$ quantities control the leakage of different instrumental systematics into the cosmological signal; consequently, smaller values of $|\tilde{h}_n|^2$ are desirable for systematics suppression.

In the case of the MeerKAT scanning strategy that was used in \cite{2020arXiv201113789W}, it is possible to obtain closer-to-optimal systematics suppression with only a few small changes. In Table~\ref{table:hn} we present the values of $|\tilde{h}_n|^2$ averaged across the map for the scan strategy used by MeerKAT. We also show the new values after supplementing the scan strategy with an additional set of scans performed at the optimal elevation for suppressing $|\tilde{h}_2|^2$, which happens to be $39^\circ$ for this field -- slightly lower than the elevations already used. The scans are performed when the target field is both rising and setting. The approach to calculate the optimal elevation to minimise a given $|\tilde{h}_n|^2$ for a particular field is presented in Appendix~\ref{section:Scanning Strategies}. Despite selecting the optimal elevation to suppress $|\tilde{h}_2|^2$ for the given field, there is only a modest improvement to the $\tilde{h}_n$ values due to the limitations that ground-based scanning places on crossing angle coverage. As such, despite being the optimal choice, this would not significantly reduce the effects of the corresponding systematics. %In Appendix~\ref{section:Scanning Strategies} we show that this extra elevation also makes a small improvement to the efficacy of the map-making approach for removing the systematics.

\begin{table}
 \caption{$|\tilde{h}_n|^2$ values averaged over the observed pixels for the MeerKAT scan strategy of \protect\cite{2020arXiv201113789W}, and an enhanced scan strategy which includes additional observations at the optimal elevation to suppress $|\tilde{h}_2|^2$ for this field. These quantities control the leakage of different instrumental systematics into the cosmological signal; smaller values of $|h_n|^2$ are desirable for systematics suppression.}
\centering
\begin{tabular}[1.8]{|c|cccc|} 
 \hline 
 & $|\tilde{h}_1|^2$ & $|\tilde{h}_2|^2$ & $|\tilde{h}_3|^2$ & $|\tilde{h}_4|^2$   \\
\hline 
MeerKAT strategy & $0.94$ & $0.78$ & $0.64$ & $0.59$\\
Enhanced strategy & $0.92$ & $0.72$ & $0.54$ & $0.50$\\
 \hline 
 \end{tabular} 
 \label{table:hn}
\end{table}

\section{Case studies}
\label{section: Spin Characterisation of Systematics}
In this section we illustrate the approach by constructing Eqs.~\ref{eq:decomp} and \ref{eq:SpinSyst} for two specific instrumental systematics.

We assume that the signal observed is solely a function of RA, Dec, and crossing angle, that maps are made at each frequency separately, and that at each frequency the observed signal does not mix with other frequencies. Both of our case studies will assume that the dish architecture and other instrumental factors lead to effective mismatches between the two polarized feeds, introducing $\psi$-dependent signals into Eq.~\ref{eq:summedsignal}.

\subsection{Polarization leakage from effective gain mismatch}
An effective mismatch in the receiver gains between the two feeds can cause polarization leakage, due to the polarization signal no longer cancelling when Stokes I maps are formed. For a simple two feed example we may model the effect of gains on the observed signal as
\begin{equation}
\begin{split}
 S(\psi,\theta,\phi) &= \frac{1}{2}((1+g^A) d_A + (1+g^B) d_B)
 \\
 %&= \frac{1}{2}\{(1+g^A)[I + Q\cos(2\psi) + U\sin(2\psi)]
 %\\
 %&+ (1+g^B)[I + Q\cos(2\psi+\pi) + U\sin(2\psi+\pi)]\}
 %\\
 &= \left(1+\frac{g^A + g^B}{2}\right)I + \left(\frac{g^A-g^B}{2}\right) \left(Q\cos(2\psi) + U\sin(2\psi)\right)
 \\
 &= \left(1+\frac{g^A + g^B}{2}\right)I + \left(\frac{g^A-g^B}{2}\right)\frac{1}{2} \left(Pe^{-2i\psi} + P^{*}e^{2i\psi}\right)
 \label{equation:GainLeakage}
\end{split}
\end{equation}
where $g^X$ is the gain offset of feed $X \in \{A,B\}$, and $P=Q+iU$. We may thus write the observed spin-0 signal as
\begin{equation}
    {}_{0}\tilde{S}^{d} = \tilde{h}_{0}\left(1+\frac{g^A + g^B}{2}\right)I + \left(\frac{g^A-g^B}{4}\right) \left(\tilde{h}_{-2}P + \tilde{h}_{2}P^{*}\right).
\end{equation}
The first term represents the amplification of the intensity signal itself, the second term represents the leakage of the polarization signal due to any gain mismatch ($g^A-g^B$) between the feeds. The polarization leakage is typically more troublesome as it is not so easily calibrated out.

\subsection{Beam squint}
\label{section:Beam Squint}
Beam squint is a frequency-dependent offset of the beam centre from the nominal pointing centre. The squint will be different for each of the linearly polarized feeds of each antenna and will also vary in frequency \citep{2021MNRAS.502.2970A,2008A&A...486..647U,2011ITAP...59.2004H}. We model beam squint according to this definition as a frequency-dependent pointing error of each linearly polarized feed.

Assuming that the beam squint is a constant in the instrument frame, for a boresight pointing $(x,y)$, the actual pointing of feeds $A,B$ is $(x+\delta x_A, y+\delta y_A)$ and $(x+\delta x_B, y+\delta y_B)$, which will depend on the crossing angle $\psi$. The signals from a feed can be expressed in terms of the instrument frame polar offsets $\rho$ and $\chi$, the magnitude and direction of the pointing error as $\delta x_X=\rho_X \sin(\psi + \chi_X)$ and $\delta y_X=\rho_X \cos(\psi + \chi_X)$ where $X \in \{ A,B \}$ \citep[see Section 2.1 of][]{Wallisetal2016}.

As the instrument observes the sky, the observed signal of each feed is subject to a pointing error. Consequently, a further dependence on $\psi$ is imparted on the observed signal -- not only is the spin-0 intensity measured, but additional spin signals manifest due to the pointing error affecting the observation of the intensity and polarization signals. Since the intensity is spin-0 and the pointing offset imparts an $e^{i \psi}$ dependence, this causes a spin-1 leakage from the derivative of the intensity field to appear. The polarization signal is slightly more complicated as it is a spin-2 field -- in this case the pointing error causes both spin-1 and 3 signals to manifest.

Following \cite{Wallisetal2016} and \cite{2021MNRAS.501..802M}, and considering the case with an offset of magnitude $\rho_{A}$ in direction $\chi_{A}$ for feed A, and $\rho_{B}$ in direction $\chi_{B}$ for feed B, summing the two feeds shows the observed signal is contaminated as
\small
\iffalse
\begin{equation}
\begin{split}
    &S(\psi,\theta,\phi) = \frac{1}{2}(d_A + d_B)
    \\
    &= I + \frac{1}{2}\Big[\frac{1}{2}\left((\rho_Ae^{i\chi_A} +\rho_Be^{i(\chi_B + \frac{\pi}{2})}) e^{i\psi}\eth I + (\rho_Ae^{-i\chi_A} +\rho_Be^{-i(\chi_B + \frac{\pi}{2})}) e^{-i\psi}\bar{\eth} I \right)
    \\
    &-\frac{1}{4} \Big( (\rho_Ae^{i\chi_A} +\rho_Be^{i(\chi_B + \frac{3\pi}{2})}) e^{i3\psi}\eth P+ (\rho_Ae^{-i\chi_A} +\rho_Be^{-i(\chi_B - \frac{\pi}{2})}) e^{i\psi}\bar{\eth}P
    \\
    &+(\rho_Ae^{i\chi_A} +\rho_Be^{i(\chi_B - \frac{\pi}{2})}) e^{-i\psi}\eth P^*+(\rho_Ae^{-i\chi_A} +\rho_Be^{-i(\chi_B + \frac{3\pi}{2})}) e^{-3i\psi}\bar{\eth}P^* \Big)\Big],
\end{split}
\end{equation}
\fi
\begin{equation}
\begin{split}
    &S(\psi,\theta,\phi) = \frac{1}{2}(d_A + d_B) = I + \frac{1}{2}\Big[\frac{1}{2}\left(\zeta e^{i\psi} \bar{\eth} I + \zeta^* e^{-i\psi}\eth I \right)
    \\
    &-\frac{1}{4}\left( \zeta e^{i3\psi}\bar{\eth} P^*+ \zeta^* e^{i\psi}\eth P^* +\zeta e^{-i\psi}\bar{\eth} P+\zeta^* e^{-3i\psi}\eth P\right)\Big],
\end{split}
\label{equation:BSLeakage}
\end{equation}
\normalsize
where we use the spin raising operator $\eth = \frac{\partial }{\partial y}+i\frac{\partial }{\partial x}$ and its conjugate (the spin lowering operator, denoted by a bar), and where we have set $\zeta = \rho_Ae^{i\chi_A} +\rho_Be^{i\chi_B}$, which represents the total combined beam squint having summed the two linearly polarized feeds. We may thus write the observed spin-0 signal as
\begin{equation}
\begin{split}
    {}_{0}\tilde{S}^{d} &= \tilde{h}_{0} I + \frac{1}{4}\left(\tilde{h}_{1}\zeta\bar{\eth} I + \tilde{h}_{-1}\zeta^*\eth I \right)
    \\
    &-\frac{1}{8}\left(\tilde{h}_{3}\zeta\bar{\eth} P^*+ \tilde{h}_{1}\zeta^*\eth P^* +\tilde{h}_{-1}\zeta\bar{\eth} P+\tilde{h}_{-3}\zeta^*\eth P\right).
\end{split}
\end{equation}
This represents a simple model of the beam squint systematic constructed according to our formalism. We note that in the simulations presented in Section~\ref{section:Beam Squint Removal} to explore beam squint we set $\rho_B=0$ and $\chi_A=0$ such that $\zeta = \rho_A$ -- this allows the results to be presented more clearly but does not affect the bottom line conclusions of this work.

\section{Demonstration of Extended Map-Making}
\label{section:Map-Making}
Having identified the spins that are relevant for each of our two case studies in the previous section, we now show how this allows the use of map-making to remove these systematics. Given we know the spin of the systematics, in principle it is possible to solve for them along with the intensity signal using information from the scanning strategy.

The standard procedure in intensity mapping is to take the TOD and generate a map at each frequency band. These maps will ideally contain only the spin-0 intensity signal, which will include both the cosmological signal and foregrounds. Foreground removal is then applied to the output maps using a chosen component separation method \citep[e.g.][]{Wolz:2013wna, 2015MNRAS.447..400A,Olivari:2015tka,2020arXiv201002907C,2020MNRAS.499..304C,2021arXiv210512665S,2021arXiv210702267I,2021arXiv210710814S}. However foregrounds with a non-smooth frequency-dependent response introduced by the systematics prove difficult to remove using the standard approaches.

The techniques we present in this section offer a way to remove certain systematic-induced leakage at the map-making stage, isolating the spin-0 sky signal from systematic signals of other spin, e.g. spin-2 polarization contamination. The observed signal should then be closer to a pure spin-0 intensity field whose foregrounds are smooth in frequency. This would allow for the foreground removal procedures to progress without the worry of spurious signals of other spin.

\subsection{Simulations}
\begin{figure*}
  \centering
    \includegraphics[width=2\columnwidth]{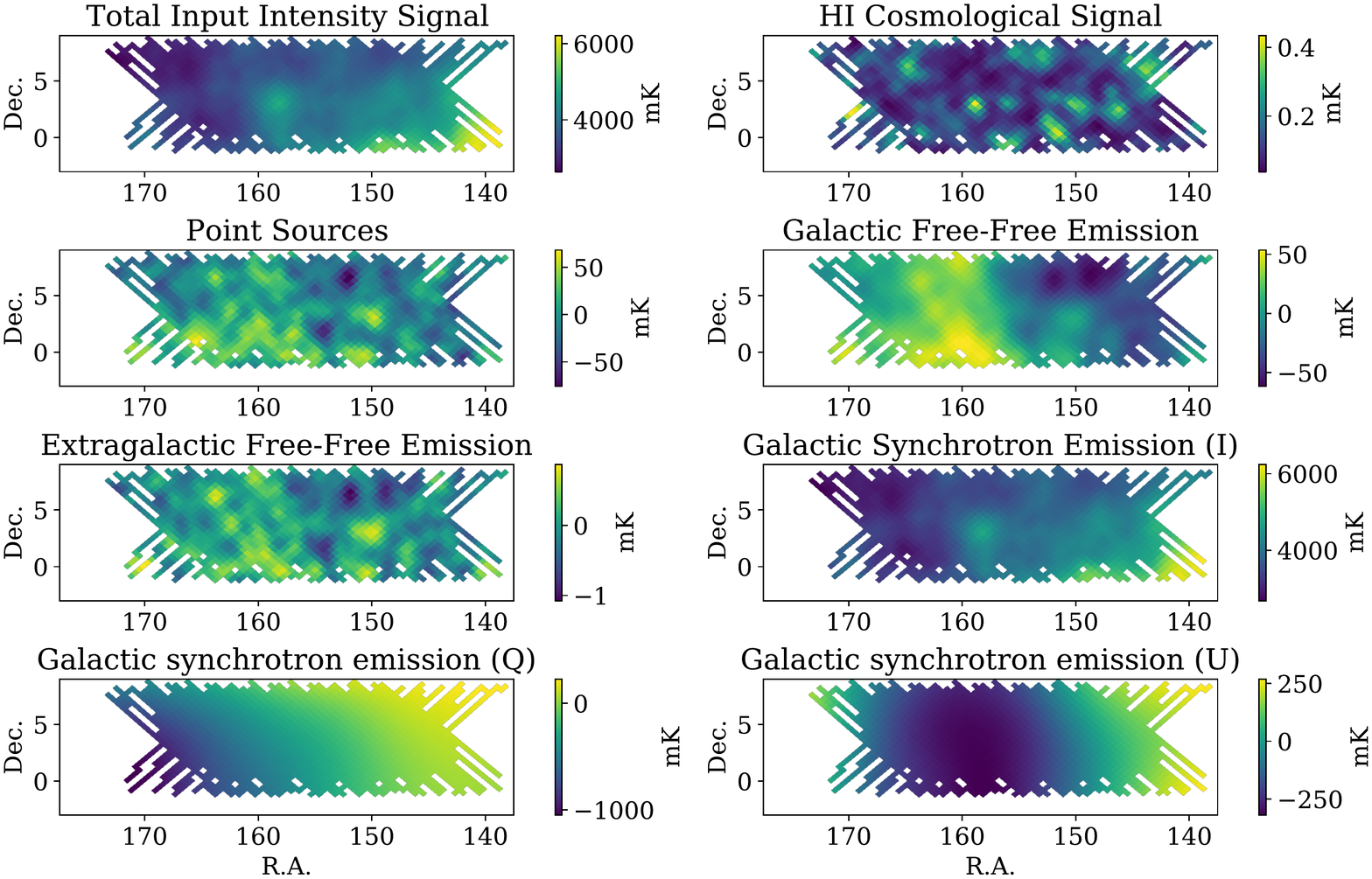}
\caption{Maps of the simulated foregrounds and HI Cosmological signal at $\sim915.5$MHz used as input to the simulations. The full sky input maps have been masked appropriately to include only the region of sky that the scanning strategy covers. The dominant foreground is the unpolarized galactic synchrotron emission, shown for Stokes $I$. There is also a large polarized component to the foregrounds from the galactic synchrotron emission, shown for Stokes $Q$ and $U$.}
\label{figure:InputSignals}
\end{figure*}

\begin{figure}
  \centering
    \includegraphics[width=1\columnwidth]{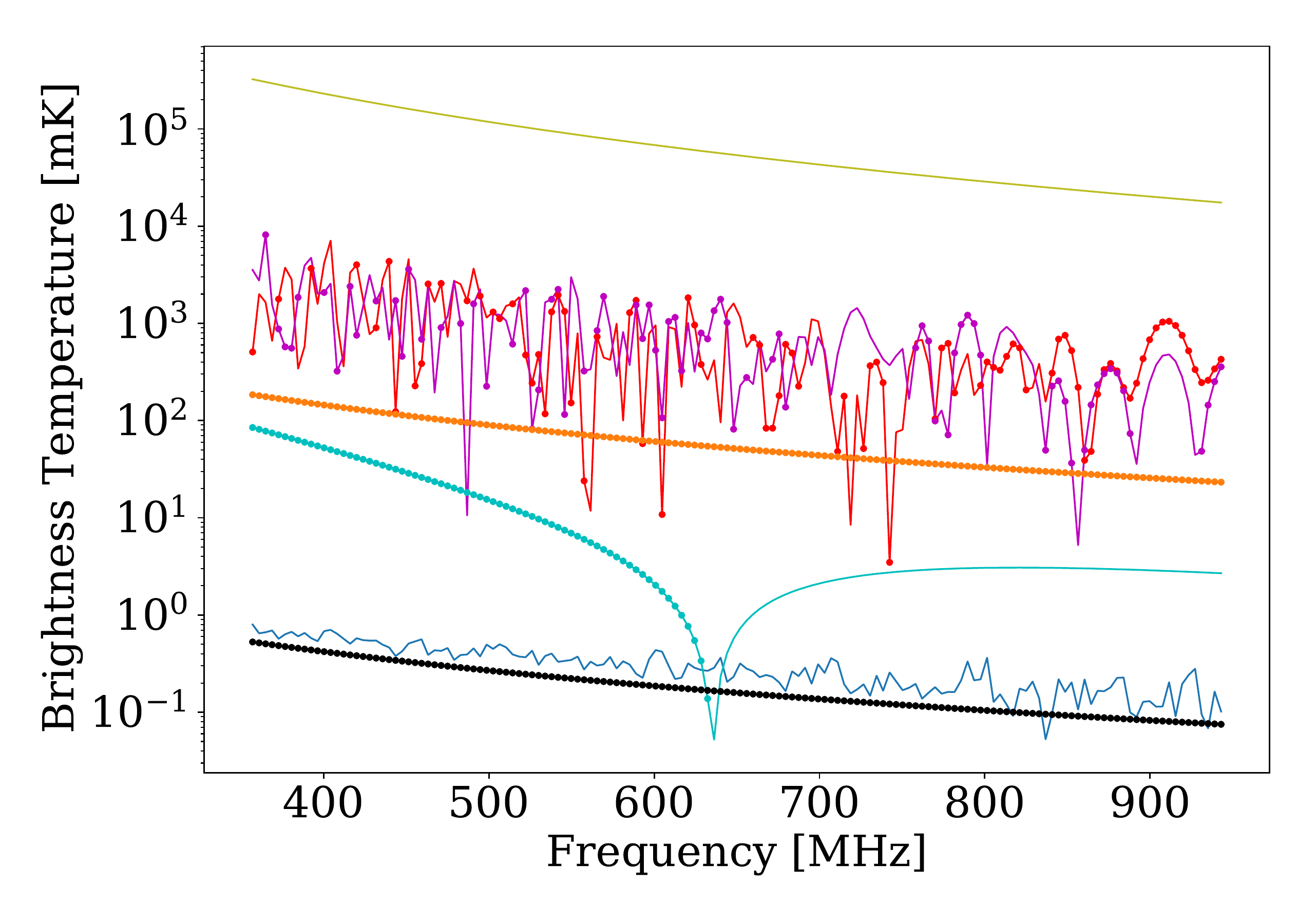}
    
    \includegraphics[width=1\columnwidth]{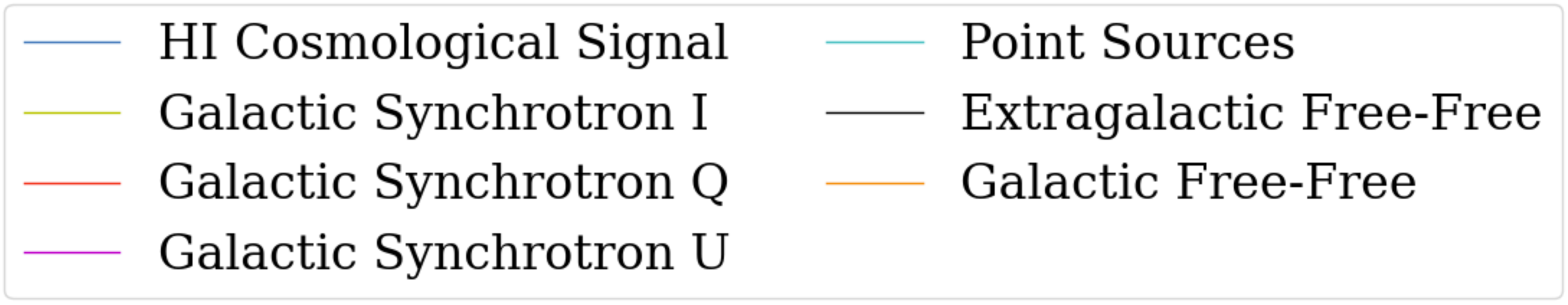}
\caption{The absolute value of the input signals as a function of frequency plotted for data from a single pixel at R.A.$=155^\circ$ and Dec.$=3^\circ$ -- roughly corresponding to the centre of the MeerKAT field that we consider. The negative data is indicated by point markers. The unpolarized foregrounds all have smooth frequency dependence -- whereas the polarized galactic synchrotron varies non-smoothly with frequency due to frequency dependent Faraday rotation.}
\label{figure:InputSignalsFrequency}
\end{figure}

In order to assess the impact of solving for multiple spin signals during map-making we produce some simple simulations as follows. We adopt the MeerKAT scanning strategy summarised in Tables 1 and 2 of \cite{2020arXiv201113789W} to generate telescope pointing data targeting the WiggleZ 11hr field. A couple of slight differences should be noted: we do not include the diode injection, and consider only a single dish. A TOD simulation is then used to sample simulated sky maps with either a beam squint or an effective gain mismatch systematic applied, which we take to be stable in time for the extent of the survey.

The simulated sky maps include the HI intensity signal, along with both polarized and unpolarized foregrounds. We generate the signals in 150 frequency bands ranging from 355-945~MHz using the {\sevensize CRIME}\footnote{\url{http://intensitymapping.physics.ox.ac.uk/CRIME.html}} software's ``Fast'' simulations \citep{2014MNRAS.444.3183A}. Note that this frequency range is more similar to an SKAO-MID Band 1 or MeerKAT UHF band survey than the higher-frequency MeerKAT L-band survey in \cite{2020arXiv201113789W}. We use the same input cosmology and foregrounds as in \cite{2014MNRAS.444.3183A}, including the cosmological HI intensity signal, along with the following foregrounds: synchrotron emission, point sources, extra-galactic and galactic free-free emission, and polarized synchrotron emission. An example of these is shown for a single frequency channel in Fig.~\ref{figure:InputSignals}.

In Fig.~\ref{figure:InputSignalsFrequency} we show the input signals as a function of frequency for data from a single pixel at R.A.$=155^\circ$ and Dec.$=3^\circ$ -- roughly corresponding to the centre of the field considered. The unpolarized foregrounds have smooth frequency dependence, whereas the polarized galactic synchrotron varies non-smoothly with frequency. This non-smooth variation stems from the frequency-dependent Faraday rotation. \cite{2014MNRAS.444.3183A} include this by using the statistical properties of the synchrotron emission in the line of sight space of Faraday depths. The angular structure of the field is based on the unpolarized counterpart, and the structure in frequency is injected according to a correlation length in Faraday depth space which is derived from the complex models of {\sevensize Hammurabi} \citep{2009A&A...495..697W} (see \cite{2014MNRAS.444.3183A} for more detail on the foreground simulations).

We use {\sevensize HEALPY} \citep{Zonca2019} maps with a resolution of {\tt NSIDE=128} which we smooth using a $1^\circ$ Gaussian beam, the width corresponding to the size of the MeerKAT primary beam \citep{2020arXiv201113789W}. For simplicity, the beam model we use is independent of frequency, but note that this should not significantly affect the bottom line results since the map-making is performed separately for each frequency band. To aid in the proof of concept, we implement a noiseless setup, allowing for a clearer elucidation of the effects the different map-making approaches have on the systematics. We note that the inclusion of noise in map-making is a well-understood process \citep[e.g.][]{1997ApJ...480L..87T}, and its absence should not affect the bottom line results of the techniques we explore.

We apply a simple binned map-making procedure to the TOD to derive our observed maps, treating each frequency band separately. A spin-0 map-making approach is applied as an example of that used as standard in intensity mapping, which amounts to simply averaging the signal within each pixel. In addition we apply an extended map-making approach solving for both the spin-0 signal and further spin-dependent signals chosen to match the spin of the systematic.\footnote{We avoid complications such as destriping for the moment; however these types of techniques are ``orthogonal'' to the processes we are advocating -- they are applied directly to the timestream, occurring prior to the map-making step in the analysis process, and as such could be included without detriment \citep{2010MNRAS.407.1387S}.}

Solving for additional signals of different spin makes the map-making slightly more complicated, requiring a matrix inversion based on the crossing angles to disentangle the signals (see Eq.~\ref{eq:Spin 2 3x3 Map-making} below). This matrix inversion is well behaved for a sufficiently redundant scan strategy, however less well sampled pixels may result in issues with the inversion \citep{2003astro.ph.10787H,2009A&A...506.1511K,2021MNRAS.501..802M}. We thus impose a data cut to remove pixels where the inversion is found to be ill-conditioned. This cut comes at the expense of losing some of the observed field, the pixels that tend to have the most issues being those at the field edges which are visited relatively few times, and with few distinct crossing angles. This should be less of an issue for a larger scale survey, as the scanning strategies will likely contain further redundancies -- i.e. further hits at more crossing angles. Consequently in the case of the beam squint simulations we also incorporate an additional elevation of $39^{\circ}$ into the scanning strategy for one rising and one setting scan in order that we retain more survey area at map-making -- see Appendix~\ref{section:Scanning Strategies} for more information.
%We apply a cut to any pixel where the map-making matrix condition number, calculated using a singular value decomposition, in {\sevensize numpy}\footnote{\url{https://numpy.org/doc/stable/reference/generated/numpy.linalg.cond.html}}, is greater than $10^{11}$.

Following this, in order to separate the HI cosmological signal from the foregrounds, we apply a blind component separation technique. These foreground removal techniques, for the most part, use the high dynamic range between the cosmological and foreground signals to derive spectral templates, e.g. by performing an eigenvalue decomposition on the frequency-frequency covariance matrix of the data. After selecting a number of modes $n_s$ with the highest SNR (largest eigenvalues), these are then subtracted from the observed maps.
These modes tend to be smooth in frequency, like the dominant foreground emission. This means that any systematic contributions that are stable in time and that are also smooth in frequency will be absorbed into the foreground contribution that is removed by the component separation process. This does not apply to any systematic contributions which are non-smooth in frequency however, which should mostly survive foreground removal. In this case, other cleaning procedures are required. The technique that we demonstrate here removes these systematics {\it prior} to the usual foreground removal process, i.e. at the map-making stage.

For this study, we use the component separation software {\sevensize gmca4im}\footnote{\url{https://github.com/isab3lla/gmca4im}} \citep{2020MNRAS.499..304C}, which applies the sparsity-based Generalised Morphological Component Analysis (GMCA) algorithm of \cite{4337755} to intensity mapping. We apply this technique to the observed maps from both the spin-0 dependent map-making and the extended map-making with additional spins.

\subsubsection{Simplifications and their consequences}
\label{section: Simplifications and Consequences}
We have made a number of simplifications in the setup in order to more straightforwardly demonstrate a proof of concept of this method. Here we summarise these simplifications and comment on their possible consequences in more complex scenarios.
\begin{itemize}
    \item We use a beam model that is independent of frequency. If we were to allow a frequency dependence of the beam, this would cause some additional complications in the blind foreground cleaning process. In particular, a strong variation with frequency would cause similar difficulties to those presented by Faraday rotation, particularly at higher frequencies. We do not expect this to limit the performance of the extended map-making itself however, since the map-making is performed separately for each frequency band; the extended map-making should still remove signals of higher spin from the intensity. However, the map-making method has not been built to remove the frequency-dependent signature emanating from the beam itself, which would likely not have a well-defined spin-dependence, so this would still persist and need to be removed by deconvolution methods.
    \item We use a circular Gaussian beam model. More realistic beam models would include sidelobes, and may be elliptical to some extent. The extended map-making should still work in separating the signals of different spin, however the signature of the more complicated beam pattern would be imprinted on the maps. This could therefore cause complications in the subsequent foreground cleaning process and when performing beam deconvolution.
    \item We use a noiseless setup, but in reality there would be noise affecting the data. The inclusion of noise in map-making is a well-understood process however; in the presence of random noise, the extended map-making would still reconstruct the expected signals, but with an added contribution from (e.g.) white noise. More complicated noise signatures, such as the correlated $1/f$ noise resulting from time-varying gains, would need to be dealt with prior to map-making by filtering or destriping the timestream data \citep{1997ApJ...480L..87T,2010MNRAS.407.1387S}.
    \item We use a frequency range of $355-945$~MHz for the simulations. This choice allows us to be in the regime where the Faraday rotation-induced frequency dependence of the polarized foregrounds is highly non-smooth. This highlights the ability of the extended map-making method to remove this troublesome signal, which in turn aids the performance of the blind foreground cleaning. This frequency range does not match the MeerKAT L-band survey in \cite{2020arXiv201113789W}, whose scan strategy we adopt in the simulations. Nonetheless, the scanning strategy used would not be unrealistic for the frequency range adopted, which is similar to an SKAO-MID Band 1 or MeerKAT UHF band survey. If we were to work at a higher frequency range closer to e.g. MeerKLASS \citep{2017arXiv170906099S}, the signature from Faraday rotation would be more smoothly oscillatory in nature, as is evident from Fig.~\ref{figure:InputSignalsFrequency}. Since it would have a smoother frequency dependence and be less noise-like in nature, existing blind foreground cleaning algorithms would likely fair better than at the frequencies we present. We expect, however, that the extended map-making should still be beneficial, helping to reduce the number of sources that need to be included in the cleaning algorithms.
\end{itemize}

\subsection{Removing polarization leakage}
\label{section:Polarization Removal}
The non-smooth frequency dependence of the polarization leakage in HI surveys is a result of Faraday rotation and other line of sight mixing effects. It has been noted that current methods that deal with polarization leakage suffer due to the Faraday rotation effect, and when they are applied can also lead to the loss of the cosmological signal itself \citep{2020arXiv201002907C}. The method we present provides a way to disentangle the leaked polarization signal while incurring less loss of the HI signal itself.

We use a map-making approach that removes the polarization leakage for each frequency individually, meaning the frequency dependence of the Faraday rotation does not affect the ability of the method to remove the polarization signal and safely extract the intensity signal. Different positions on the sky experience different amounts of Faraday rotation, and the angular structure of this variation can be varied in simulations depending on the smoothing scale used. The ``Fast'' simulations of \cite{2014MNRAS.444.3183A} used a $5^{\circ}$ smoothing of the Faraday depth maps of \cite{2012A&A...542A..93O}. Since we use a map of {\tt NSIDE=128} in the simulations, this means the Faraday rotation will be the same throughout the extent of each pixel. Provided this representation of Faraday rotation is sufficiently representative, then the method we advocate is robust to its effects -- the linear polarization will still transform as a spin-2 field regardless of this rotation, and our pixel by pixel approach to map-making means that the variation in the Faraday rotation across the sky will not affect the ability of the map-making to retrieve the signal.
%It is also worth pointing out that if circular polarization contamination is anticipated, the map-making process could also be fairly simply extended to solve for the Stokes V signal which would remove the contamination from the observed intensity signal.

In the case of effective gain mismatch, there will be an amplification of the intensity signal usually dealt with by careful calibration. However there will also be leakage of polarization due to any difference in the gain offset between the feeds as in Eq.~\ref{equation:GainLeakage}.

\begin{figure*}
  \centering
    \subfloat[Input HI cosmological signal at $\sim915.5$MHz.\label{figure:map_HI_2}]{\includegraphics[width=1\columnwidth]{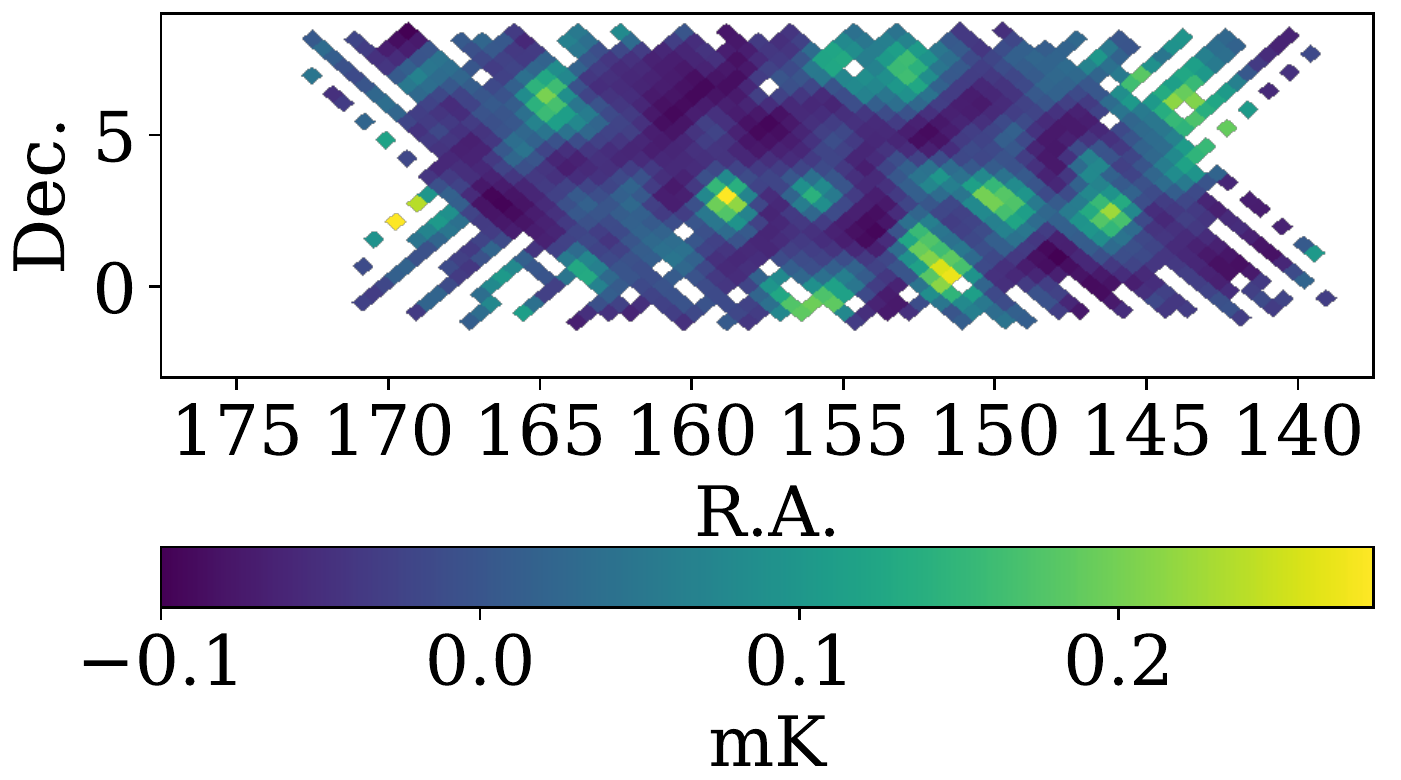}}
    \\
    \subfloat[Foreground cleaned maps at $\sim915.5$ MHz for the effective gain mismatch systematic simulation. Left Panel: Spin-0 and 2 map-making, where $n_s=5$ was used. Right Panel: Spin-0 map-making, where $n_s=6$ was used. \label{figure:map_gain}]{\includegraphics[width=2\columnwidth]{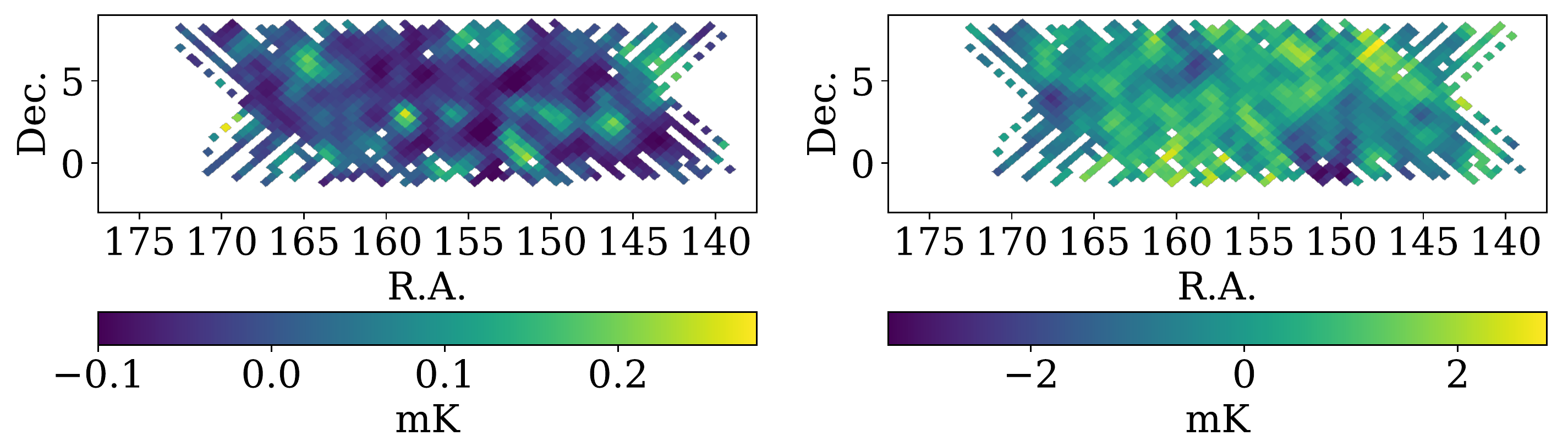}}
\caption{Comparing panels \ref{figure:map_HI_2} and \ref{figure:map_gain} we see that when an effective gain mismatch is present, the retrieved signal is dominated by polarization leakage when applying standard spin-0 map-making. However in the extended spin-0 and 2 map-making case we see that the retrieved signal is no longer contaminated by polarization as the map-making has removed the leakage prior to the foreground extraction -- allowing the HI signal to be retrieved successfully.}
\label{figure:map_g}
\end{figure*}

For the case of simple binned map-making, since we have summed the two feeds, this means in principle we can directly map-make for the intensity signal as
\begin{equation}
    I = \langle d_j \rangle
    \label{equation:ImapMake}
\end{equation}
where the angle brackets $\langle \rangle$ denote an average over the $j$ measurements in a pixel, and $d_j$ is the TOD signal summed from the two linearly polarized feeds. However this will result in the spin-2 polarization signal, present due to the gain mismatch, leaking into the observed intensity signal. In order to avoid this we may include spin-2 signals into map-making as
\begin{equation}
\begin{pmatrix}I\\Q\\U\end{pmatrix}
=
R_2^{-1}
\begin{pmatrix}\langle d_{j} \rangle\\\langle d_{j}\cos(2\psi_{j}) \rangle\\\langle d_{j}\sin(2\psi_{j}) \rangle\end{pmatrix},
\label{eq:Spin 2 3x3 Map-making}
\end{equation}
where
\begin{equation}
R_2 \equiv \begin{pmatrix}1&\langle \cos(2\psi_{j}) \rangle&\langle \sin(2\psi_{j}) \rangle\\\langle \cos(2\psi_{j}) \rangle&\langle \cos^{2}(2\psi_{j}) \rangle&\langle \cos(2\psi_{j})\sin(2\psi_{j}) \rangle\\\langle \sin(2\psi_{j}) \rangle&\langle \sin(2\psi_{j})\cos(2\psi_{j}) \rangle&\langle \sin^{2}(2\psi_{j}) \rangle\end{pmatrix} \nonumber\text{.}
\end{equation}
Since we know that the gain-leaked polarization is spin-2, this means that solving for a spin-2 signal in addition to the spin-0 intensity at map-making will disentangle the two and remove the leakage. We should note here that this technique is not specific to gain-leaked polarization; it would in fact disentangle any direct leakage of the spin-2 polarization signal regardless of the systematic sourcing it, provided it is stable for the time-span of the data included in the map-making.
We also note that this map-making process also reconstructs the spin-2 signal itself. As such one could also plot maps of the systematic signal, in this case polarization leakage.

To demonstrate the ability of map-making to remove the contaminating signal we use a simulation with a 1\% effective gain mismatch ($g^{A}-g^{B} = 0.01$). In order to concentrate on the polarization leakage we have also assumed that we can perfectly calibrate out the intensity amplification due to the gain systematic.

In principle the effective gain mismatch itself could vary with frequency, however we choose not to implement this in the simulations as its introduction would not alter the bottom line results; the systematic already has non-smooth frequency dependence due to the leakage of the polarized synchrotron foreground into the observed intensity signal and as such would already escape the standard component separation procedures.

Additionally, given that gain systematics are subject to drift over time, the effective gain mismatch ($g^{A}-g^{B}$) would likely also vary with time. A time variation imparts additional structure on the leaked signal, making it lose its pure spin dependence. Applying the extended map-making would consequently fail to remove the contamination. Provided the drift in time is slow enough, it may be possible to apply the map-making to TOD split into shorter timescales over which the systematic is stable in order to remove the contamination. This would require sufficient crossing angle coverage in each pixel over the shorter time frame, so that the map-making remains well-conditioned. This may be difficult given the limits that ground-based observatories place on scanning strategies, necessitating other methods to be used that are capable of dealing with a time-varying signal \citep{2021arXiv210202284T}.

\subsubsection{Effect on foreground removal}
\begin{figure}
  \centering
    \includegraphics[width=1.1\columnwidth]{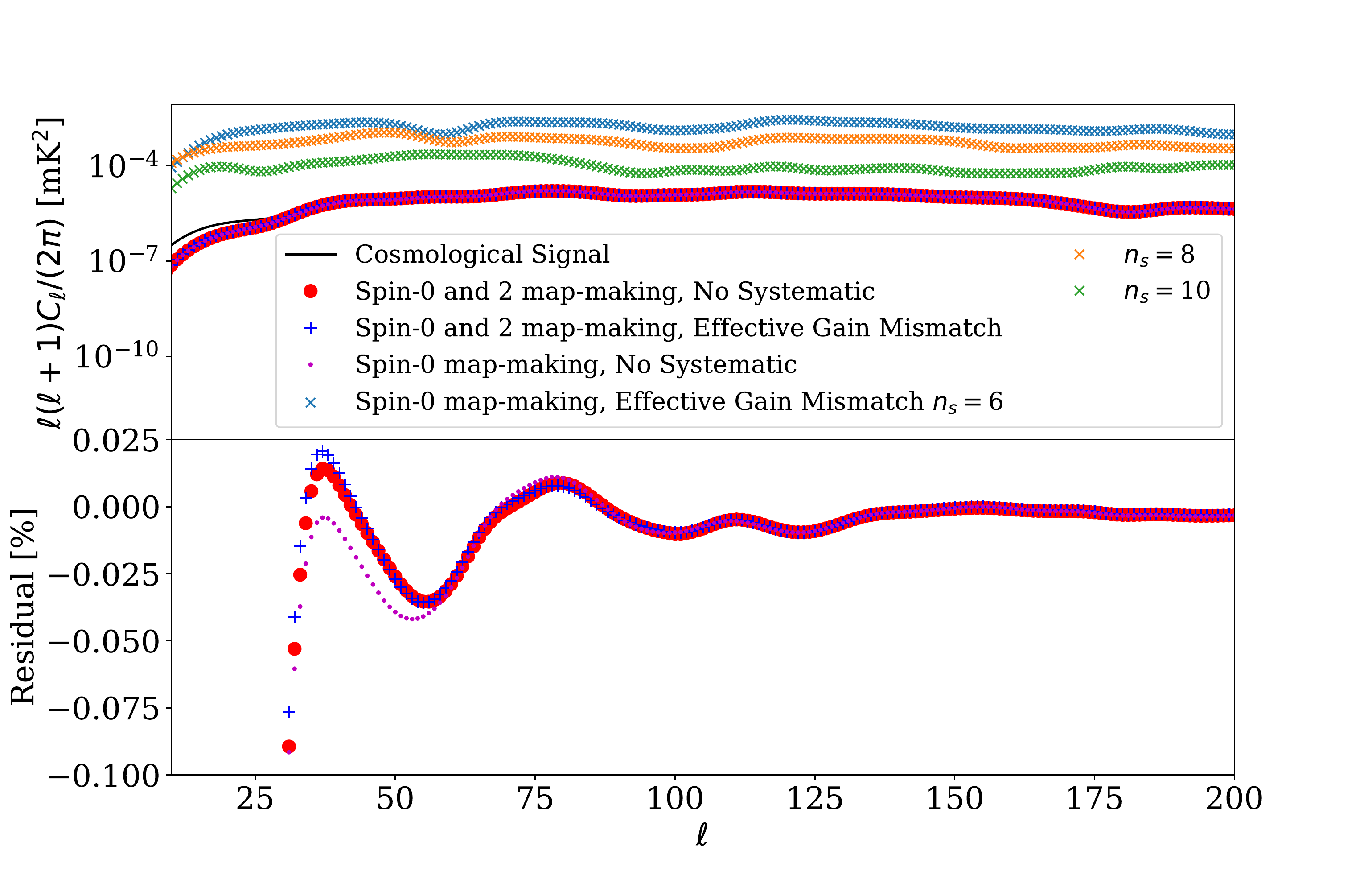}
\caption{Foreground-cleaned angular power spectra at $\sim915.5$MHz. We see that the spectra output in the absence of systematics (red circles and pink dots) matches the input HI cosmological signal well for both types of map-making. The slight disagreement at low-$\ell$ is a consequence of some loss of the cosmological signal at large angular scales during the foreground removal process. An effective gain mismatch systematic leaks polarization into the observed signal, so when implementing the standard spin-0-only map-making, this results in a large contamination to the retrieved HI angular power spectrum. This is shown as cyan, orange, and green crosses for $n_s=6$, 8, and 10 respectively; the cleaning does suppress the signal more for larger $n_s$. However, in the case where we apply the extended map-making process, solving for both spin-0 and spin-2 signals, we see that the contamination is almost completely removed (blue +s), and we have only needed to use $n_s=5$. The percentage residuals, taken between the cleaned data and the input cosmological signal and plotted in the lower panel, demonstrate that the extended map-making reduces the contamination level to be comparable to the cases with no systematics.}
\label{figure:c_ell_g}
\end{figure}

Fig.~\ref{figure:map_g} shows the output intensity maps, from the simulation including polarization leakage via effective gain mismatch, after the application of the blind foreground removal procedure where we set the number of `sources' to $n_s=5$ for the extended spin-0 and 2 map-making and $n_s=6$ for the spin-0 map-making. The different values of $n_s$ were selected to allow for a fair comparison, since the extended map-making should have reduced the number of contaminants by removing the systematic signal. All of the output maps have had the same mask applied to allow for simple comparison.
We have restricted the demonstration to a single frequency of $\sim$915.5 MHz here, however the behaviour is much the same at different frequency bands since map-making to remove the systematic is done for each frequency band individually.

Fig.~\ref{figure:map_HI_2} shows the input HI cosmological signal which ideally our output maps would exactly match. In Fig.~\ref{figure:map_gain} we see the output from the simulation where an effective gain mismatch systematic is present. The right panel shows that after blind foreground removal has been applied the retrieved signal is dominated by polarization leakage when applying standard spin-0 map-making -- the foreground removal does not remove the polarization due to its non-smooth in frequency nature. The left panel shows that when using the extended spin-0 and 2 map-making procedure the retrieved signal is no longer contaminated by polarization; the map-making has removed the leakage prior to the foreground extraction, allowing the HI signal to be retrieved successfully.

Fig.~\ref{figure:c_ell_g} and Fig.~\ref{figure:p_freq_g} show, post foreground removal, the angular power spectra at $\sim915.5$~MHz and the radial power spectra respectively. In both figures the output spectra of the simulations with no systematics are shown as red circles (spin-0 and 2 map-making) and pink dots (spin-0 map-making) and they match the input HI cosmological signal well for both cases of map-making. In Fig.~\ref{figure:c_ell_g} there is a slight disagreement at low-$\ell$ which is a consequence of some loss of the cosmological signal at large angular scales during the foreground removal process. This is evident in the percentage residuals, taken between the cleaned data and the input cosmological signal, which show a large suppression of the signal at low-$\ell$. Similarly, in Fig.~\ref{figure:p_freq_g} a loss of the cosmological signal stemming from the foreground removal procedure causes a discrepancy at low-$k_{\nu}$; the percentage residuals clearly show a suppression of the signal at $k_{\nu} \lesssim 0.4$ \citep{2020MNRAS.499..304C}.

In both Fig.~\ref{figure:c_ell_g} and Fig.~\ref{figure:p_freq_g} the cyan, orange, and green crosses show the case, where cleaning was performed with $n_s=6$, 8, and 10 respectively, where an effective gain mismatch systematic was included and we employed the standard spin-0 map-making usually used in intensity mapping surveys -- clearly the observed intensity signal is dominated by the polarization leakage in this case, even when including a large number of sources ($n_s=10$) in cleaning. The foreground removal has removed the smooth foregrounds but the non-smooth systematic leaked signal has survived the removal process. In the case where we apply the extended map-making process solving for both spin-0 and spin-2 signals, shown as blue +s, we see that the polarization contamination has been removed. The map-making has removed the polarization leakage which subsequently allowed the foreground removal process to successfully isolate the HI cosmological signal.

\subsubsection{Summary}
This case study has shown an example of how to deal with effective gain mismatch leading to leakage of the polarization into the observed intensity at the map-making step. We emphasise that this approach will work for any stable systematic that directly leaks the spin-2 polarization signal. In the standard approach of spin-0 map-making employed in intensity mapping this systematic leakage of the polarized foregrounds would heavily contaminate the observed intensity signal. We have shown here that this contamination can be controlled by a relatively simple extension to map-making.

We have demonstrated that the polarization leakage due to systematics can be removed via map-making prior to the application of the foreground removal process -- when an additional spin-2 signal is solved for at map-making, the contamination is removed from the observed intensity signal. Polarization leakage can be a significant issue for intensity mapping surveys, and consequently this extended map-making method could prove to be a useful option for removing this type of systematic contamination that can escape the standard foreground removal methods.

We are of course working in a somewhat idealised framework here, as was outlined in Section~\ref{section: Simplifications and Consequences}, however we have demonstrated the application of a simple proof of concept with encouraging results. In future work this will need to be extended to increasingly realistic scenarios. We will further discuss the limitations to this approach in Section~\ref{section:General application and limitations}.

\subsection{Removing leakage from beam squint}
\label{section:Beam Squint Removal}
In the case of beam squint a spurious signal will manifest as a leakage of the derivative of the intensity and polarization signals as in Eq.~\ref{equation:BSLeakage}.

For simple binned map-making, since we have summed the two feeds, we should be able to directly map-make for the intensity signal according to Eq.~\ref{equation:ImapMake}. However this will result in the contamination of the spin-0 intensity signal by spin-1 and spin-3 signals stemming from the beam squint systematic. We can avoid this by the inclusion of spin-1 and 3 signals into map-making as

\begin{strip}
\hrulefill
\begin{equation}
\begin{pmatrix}I\\Z^Q_{1}\\Z^U_{1}\\Z^Q_{3}\\Z^U_{3}\end{pmatrix}
=
\begin{pmatrix}
1&\langle \cos(\psi_{i}) \rangle&\langle \sin(\psi_{i}) \rangle&\langle \cos(3\psi_{i}) \rangle&\langle \sin(3\psi_{i}) \rangle
\\\langle \cos(\psi_{i}) \rangle&\langle \cos^{2}(\psi_{i}) \rangle&\langle \cos(\psi_{i})\sin(\psi_{i}) \rangle&\langle \cos(\psi_{i})\cos(3\psi_{i}) \rangle&\langle \cos(\psi_{i})\sin(3\psi_{i}) \rangle
\\\langle \sin(\psi_{i}) \rangle&\langle \sin(\psi_{i})\cos(\psi_{i}) \rangle&\langle \sin^{2}(\psi_{i}) \rangle&\langle \sin(\psi_{i})\cos(3\psi_{i}) \rangle&\langle \sin(\psi_{i})\sin(3\psi_{i}) \rangle
\\\langle \cos(3\psi_{i}) \rangle&\langle \cos(3\psi_{i})\cos(\psi_{i}) \rangle&\langle \cos(3\psi_{i})\sin(\psi_{i}) \rangle&\langle \cos^{2}(3\psi_{i}) \rangle&\langle \cos(3\psi_{i})\sin(3\psi_{i}) \rangle
\\\langle \sin(3\psi_{i}) \rangle&\langle \sin(3\psi_{i})\cos(\psi_{i}) \rangle&\langle \sin(3\psi_{i})\sin(\psi_{i}) \rangle&\langle \sin(3\psi_{i})\cos(3\psi_{i}) \rangle&\langle \sin^{2}(3\psi_{i}) \rangle
\end{pmatrix}^{-1}
\begin{pmatrix}\langle d_{i} \rangle
\\\langle d_{i}\cos(\psi_{i}) \rangle
\\\langle d_{i}\sin(\psi_{i}) \rangle
\\\langle d_{i}\cos(3\psi_{i}) \rangle
\\\langle d_{i}\sin(3\psi_{i}) \rangle\end{pmatrix},
\label{eq:5x5mapmaking}
\end{equation}
\hrulefill
\end{strip}

\noindent
where $(Z^Q_{1}+iZ^U_{1})$ and $(Z^Q_{3}+iZ^U_{3})$ are spin-1 and spin-3 fields. Solving for spin-1 and 3 signals in addition to the spin-0 intensity at map-making will disentangle them and remove the leakage from the observed signal. We should also note that in this map-making process we will also reconstruct and study the spin-1 and 3 signals themselves. %As such we could also plot maps of the systematic signal, this could be desirable to consider the level of their contribution in comparison to what is expected for a given experiment.

\begin{figure}
  \centering
    \includegraphics[width=1.1\columnwidth]{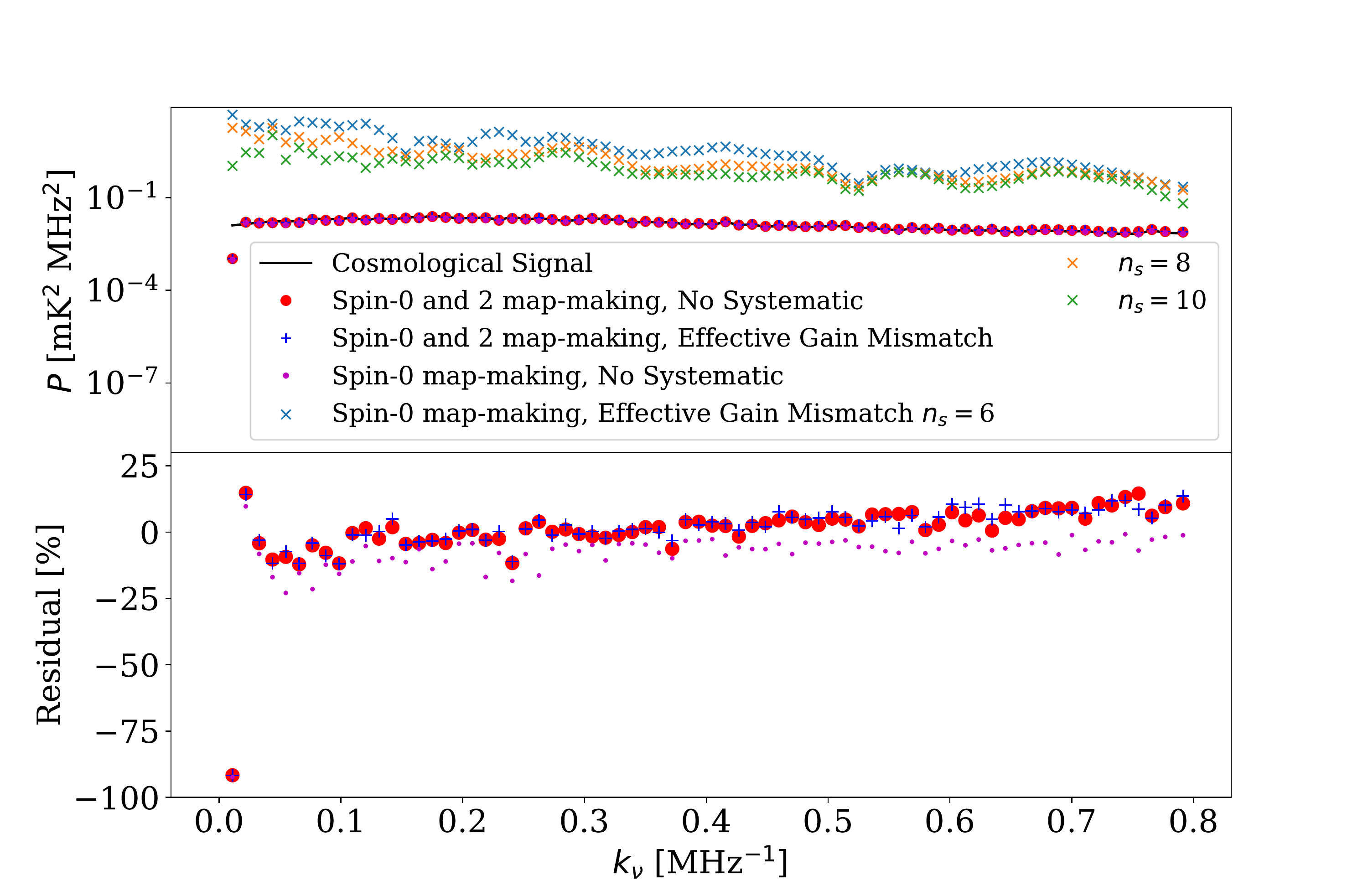}
\caption{Foreground cleaned radial power spectra. The spectra output in the absence of systematics (red circles and pink dots) matches the input HI cosmological signal well for both cases of map-making. The percentage residuals in the lower panel, taken between the cleaned data and the input cosmological signal, highlight the discrepancy at low $k_{\nu}$ ($\lesssim 0.4$ MHz$^{-1}$). This is a consequence of some loss of the cosmological signal during the foreground removal procedure. Since the effective gain mismatch systematic leaks polarization into the observed signal, when implementing the standard spin-0 map-making used in intensity mapping there is a large contamination to the retrieved HI signal, which is shown as cyan, orange, and green crosses for $n_s=6$, 8, and 10 respectively. However, when implementing the extended map-making process including both spin-0 and 2 signals, we see that the contamination is removed (blue +s), where we have used $n_s=5$. The percentage residuals plotted in the lower panel again demonstrate that the extended map-making reduces the contamination to be comparable to the cases with no systematics.}
\label{figure:p_freq_g}
\end{figure}

This approach requires a scanning strategy with sufficient redundancies such that the map-making is well behaved. As before, in our examples we impose a cut on the data when the map-making matrix is ill-conditioned to avoid issues. This results in loss of coverage of certain parts of the field however, as is evident from Fig.~\ref{figure:map_bs}. The ill-conditioning of the matrix roughly scales with the number of signals being solved for; solving for several signals simultaneously necessitates careful scan strategy design to avoid a large loss of survey area. We comment further on this in Appendix~\ref{section:Scanning Strategies}.

Fig.~\ref{figure:dI} shows that $\eth I \gg \eth P$, and as such the systematic leakage will likely be dominated by the spin-1 signal. If that is the case, it may be possible to remove the bulk of the contamination through a relatively simple inclusion of only one additional spin-1 signal at map-making as
\begin{equation}
\begin{pmatrix}I\\Z^Q_{1}\\Z^U_{1}\end{pmatrix}
=
R_1^{-1}
\begin{pmatrix}\langle d_{j} \rangle\\\langle d_{j}\cos(1\psi_{j}) \rangle\\\langle d_{j}\sin(1\psi_{j}) \rangle\end{pmatrix} \text{,}
\label{eq:Spin 1 3x3 Map-making}
\end{equation}
where
\begin{equation}
R_1 =
\begin{pmatrix}1&\langle \cos(1\psi_{j}) \rangle&\langle \sin(1\psi_{j}) \rangle\\\langle \cos(1\psi_{j}) \rangle&\langle \cos^{2}(1\psi_{j}) \rangle&\langle \cos(1\psi_{j})\sin(1\psi_{j}) \rangle\\\langle \sin(1\psi_{j}) \rangle&\langle \sin(1\psi_{j})\cos(1\psi_{j}) \rangle&\langle \sin^{2}(1\psi_{j}) \rangle\end{pmatrix} \nonumber\text{.}
\end{equation}
This simpler approach would allow for the retention of much more survey area than the extended approach of Eq.~\ref{eq:5x5mapmaking}.

\subsubsection{Smoothness of the leakage}
\label{section:smoothness}
As mentioned in Section~\ref{section:Polarization Removal}, a foreground removal procedure is applied after the map-making process. This makes use of the fact that the unpolarized foregrounds are bright and have a smooth spectral structure to separate them from the HI cosmological signal. Nevertheless, the systematic leakage could still persist after foreground removal, as systematics can introduce a non-smooth frequency dependence.

The beam squint systematic is associated with deviations from the ideal parabolic shape of a dish, leading to an effective misalignment of the observed signal. This leads to a frequency-dependent effect that also depends on instrumental parameters, as described in \cite[e.g.][]{2008A&A...486..647U,2011ITAP...59.2004H}. This effect is usually expected to be smooth in frequency, but will vary between the two polarized feeds \citep{2021MNRAS.502.2970A}. Under some circumstances, deformations in the dish architecture can lead to a non-smooth in frequency effect however \citep{2008A&A...486..647U,2011ITAP...59.2004H}.

\begin{figure}
  \centering
    \subfloat[A smooth in frequency beam squint systematic, plotted for data from a single pixel at R.A.$=155^\circ$ and Dec.$=3^\circ$. Upper Panel: $\zeta \eth I$ vs frequency. Centre Panel: $\zeta \eth P$ vs frequency. Lower Panel: $\zeta$ vs frequency.\label{figure:dIvsfrhosmooth}]{\includegraphics[width=\columnwidth]{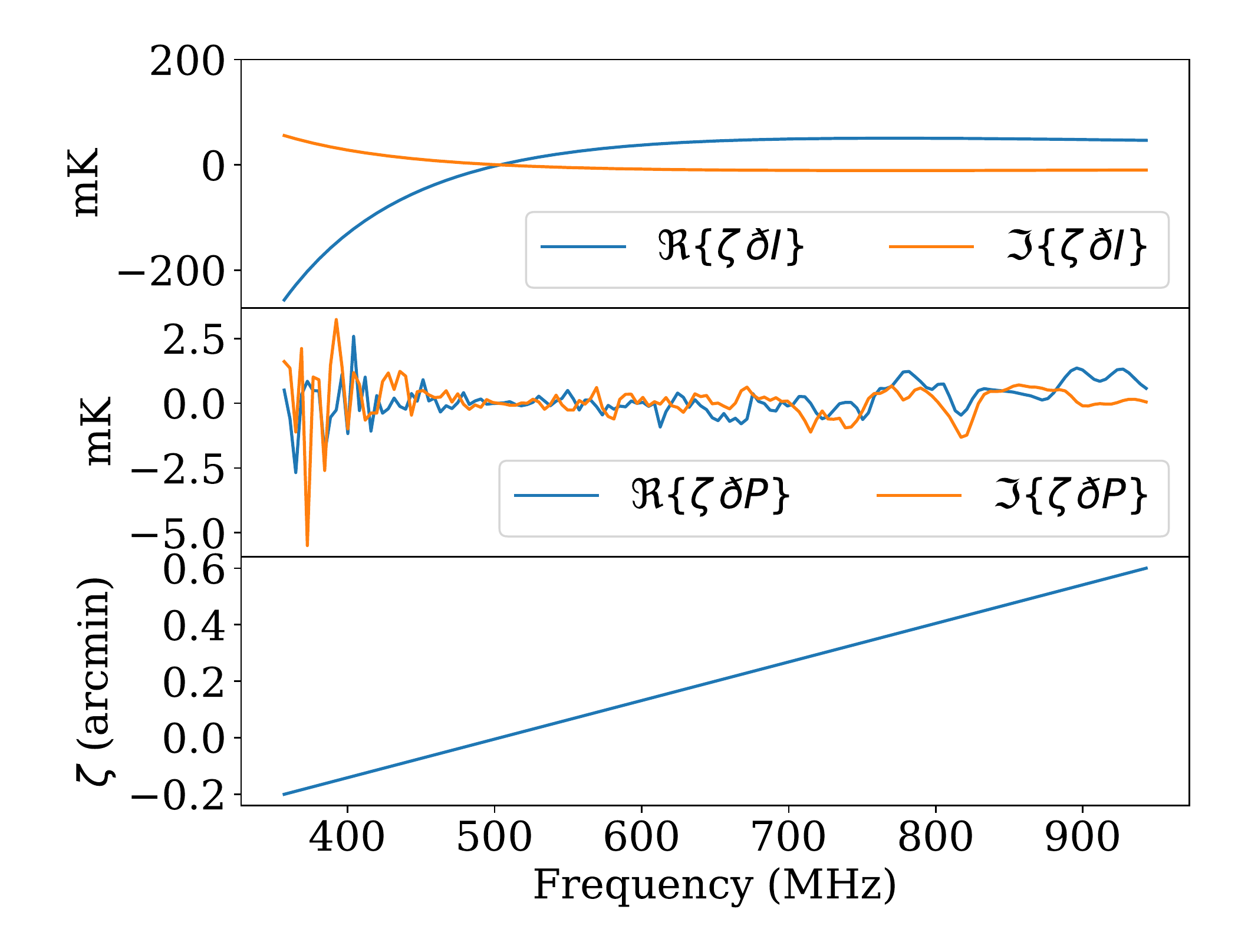}}
    \\
    \subfloat[A non-smooth in frequency beam squint systematic, plotted for data from a single pixel at R.A.$=155^\circ$ and Dec.$=3^\circ$. Upper Panel: $\zeta \eth I$ vs frequency. Centre Panel: $\zeta \eth P$ vs frequency. Lower Panel: $\zeta$ vs frequency.\label{figure:dIvsfrhorandom}]{\includegraphics[width=\columnwidth]{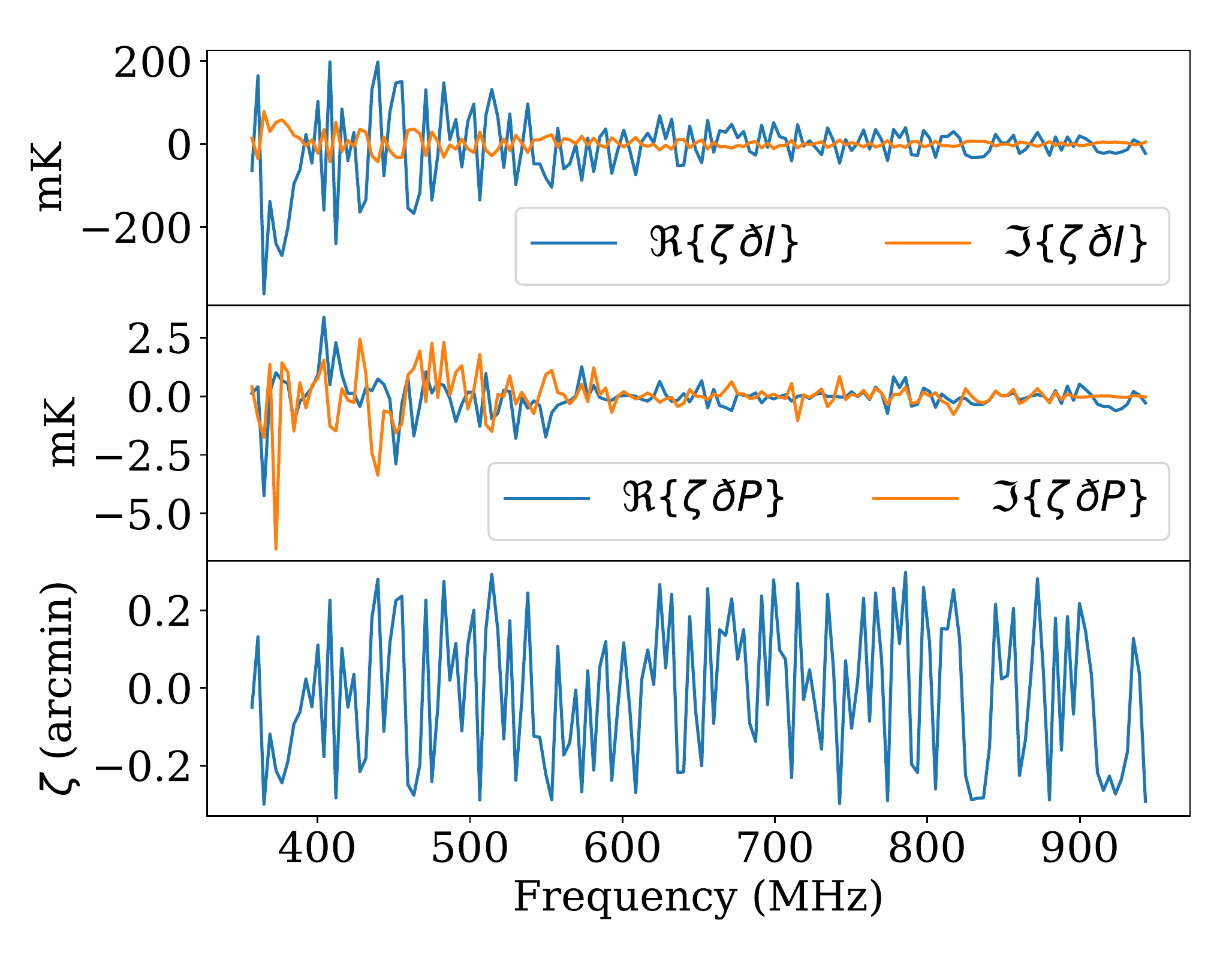}}
\caption{The $\eth I$ foreground signal is smooth in frequency and as such provided that the beam squint scaling $\zeta$ is constant or smooth (as in panel \ref{figure:dIvsfrhosmooth}) in frequency the leakage of $\eth I$ will remain smooth and possible to extract with standard blind foreground removal techniques. However $\eth P$ is non-smooth in frequency and as such even with a $\zeta$ that is constant or smooth in frequency (as in panel \ref{figure:dIvsfrhosmooth}) the leakage of this signal may evade usual foreground removal techniques. Furthermore if, as in panel \ref{figure:dIvsfrhorandom}, $\zeta$ varies non-smoothly in frequency the leaked $\eth I$ signal will no longer vary smoothly in frequency -- as such becoming inaccessible to standard blind foreground removal techniques.}
\label{figure:dI}
\end{figure}

\begin{figure*}
  \centering
    \subfloat[Input HI cosmological signal at $\sim915.5$MHz.\label{figure:map_HI_1}]{\includegraphics[width=\columnwidth]{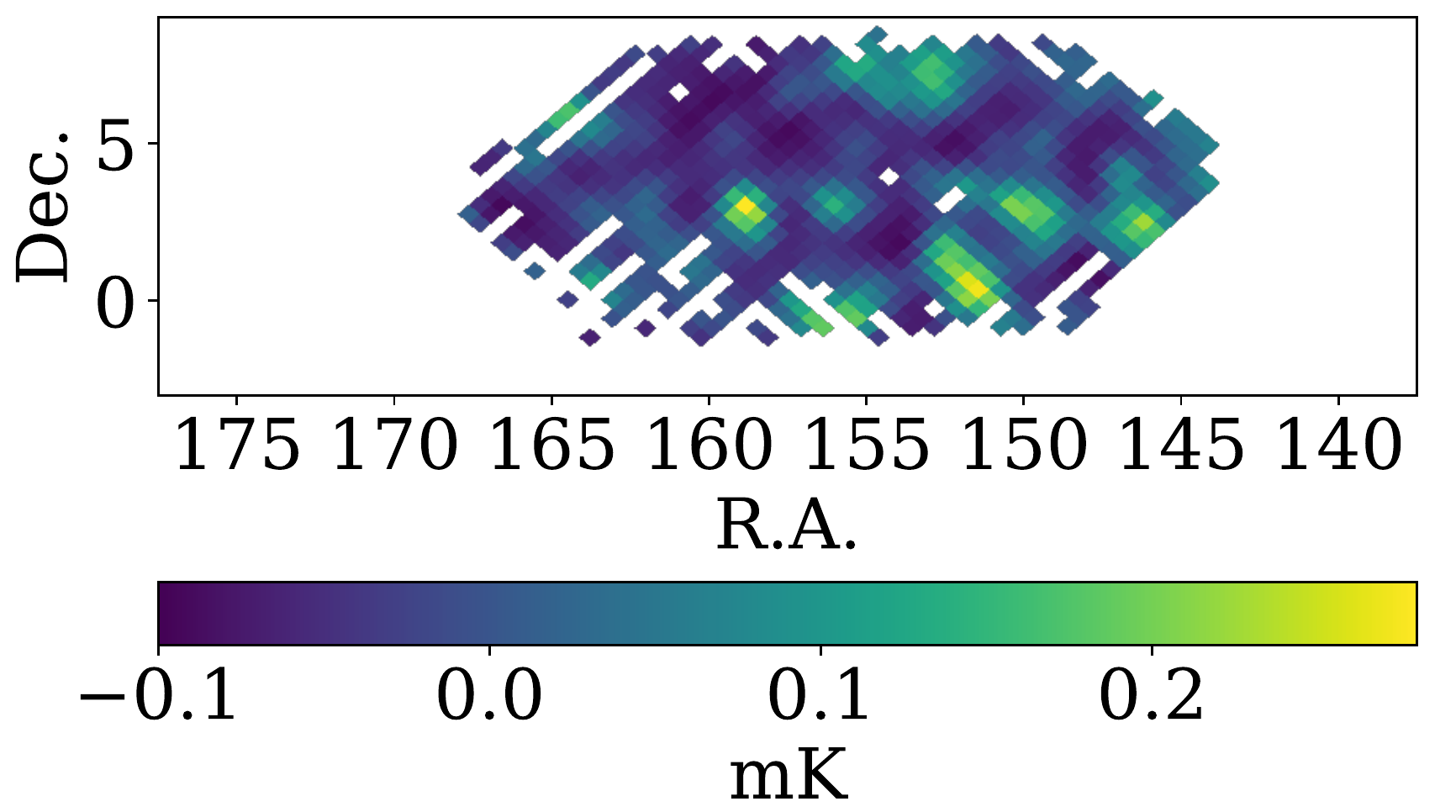}}
    \\
    \subfloat[Random beam squint systematic simulation. \label{figure:map_random}]{\includegraphics[width=2\columnwidth]{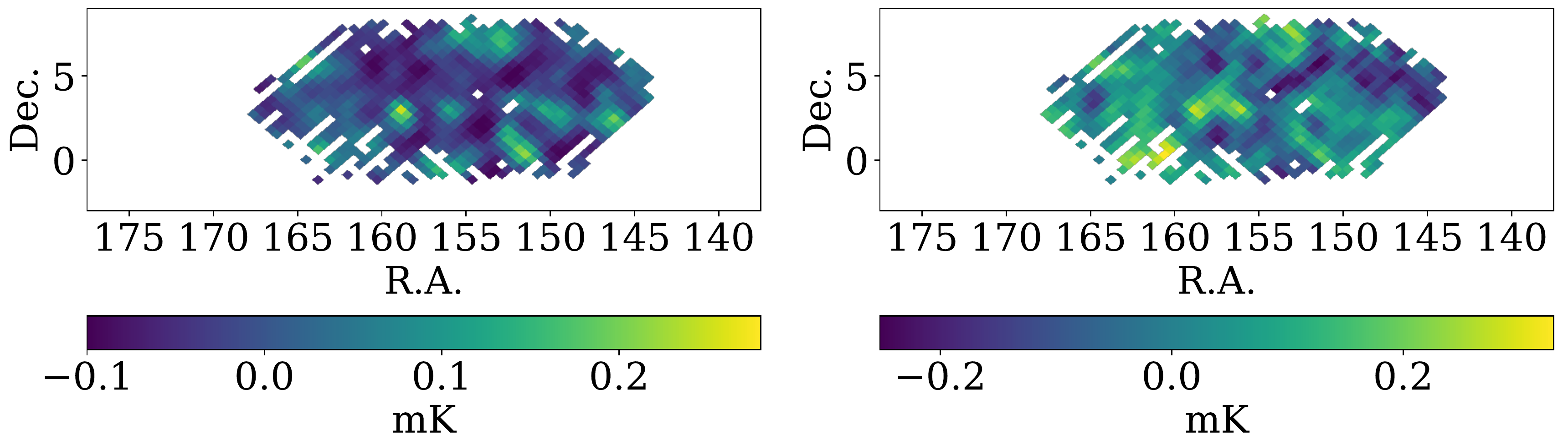}}
    \\
    \subfloat[Linear beam squint systematic simulation.\label{figure:map_linear}]{\includegraphics[width=2\columnwidth]{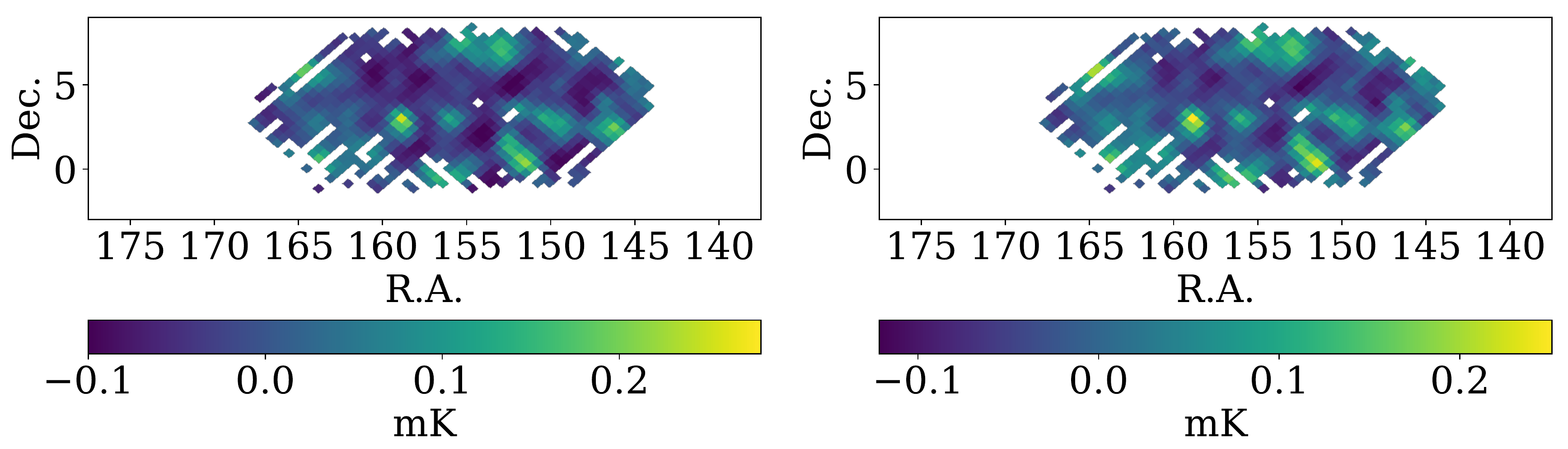}}
\caption{\ref{figure:map_random} and \ref{figure:map_linear} show foreground cleaned maps at $\sim915.5$MHz. Left Panels: Spin-0, 1 and 3 map-making with $n_s=5$ used for the cleaning. Right Panels: Spin-0 map-making with $n_s=6$ used for the cleaning. Comparing panels \ref{figure:map_HI_1} and \ref{figure:map_random} we see that for a beam squint non-smooth in frequency in the spin-0 map-making case the retrieved signal is dominated by the beam squint leakage -- however when applying the extended spin-0, 1, and 3 map-making procedure the leakage is removed prior to the foreground extraction and the HI signal is retrieved successfully. In contrast panel \ref{figure:map_linear} shows that if the systematic is smooth in frequency the foreground removal will deal almost equally well with the contamination for both types of map-making -- retrieving the HI signal successfully with only a slight difference visible in the spin-0 case due to the enduring small $\eth P$ leakage.}
\label{figure:map_bs}
\end{figure*}

Part of the beam squint systematic causes leakage of $\eth I$ into the observed intensity. Since the unpolarized foregrounds are smooth in frequency, in principle the foreground contribution to $\eth I$ is smooth in frequency also. As such, provided that the beam squint scaling $\zeta$ is constant or smooth in frequency, as illustrated in Fig.~\ref{figure:dIvsfrhosmooth}, the leakage will be possible to extract with standard blind foreground removal techniques. However if, as in Fig.~\ref{figure:dIvsfrhorandom}, $\zeta$ varies non-smoothly in frequency, the leakage of $\eth I$ will no longer vary smoothly in frequency, making it difficult to cleanly disentangle from the cosmological HI signal using standard blind foreground removal techniques.

In contrast to the $\eth I$ contamination, the contamination from $\eth P$ would be non-smooth with frequency, since the polarized foregrounds are non-smooth in frequency. As such, even in the case where the beam squint scaling $\zeta$ is constant or smooth in frequency, as seen in Fig.~\ref{figure:dIvsfrhosmooth}, the $\eth P$ component of the leakage would still evade the standard blind foreground removal techniques. In this case one could choose to remove the leakage through the inclusion of additional spins at map-making. However since $\eth P$ is a much weaker signal than $\eth I$, the leakage will be less drastic in this scenario, which means that standard foreground removal alone may suffice.

For completeness we therefore simulate both the case of a smooth in frequency squint, and the ostensibly less common scenario of a non-smooth in frequency squint to show that our approach can deal with either case. In order to investigate these cases, we run two sets of simulations:
\begin{itemize}
    \item A beam squint varying linearly ($\zeta=-0.2'$ to $0.6'$) across the 150 frequency bands. This produces a smooth frequency dependence of the $\eth I$ leakage along with a non-smooth $\eth P$ component (see Fig.~\ref{figure:dIvsfrhosmooth}).
    \item A beam squint randomly varying in frequency, sampled from a uniform distribution between $\zeta =-0.3'$ and $0.3'$. This produces non-smooth frequency dependence of the $\eth I$ and $\eth P$ leakage (see Fig.~\ref{figure:dIvsfrhorandom}).
\end{itemize}
These representative offsets $\zeta$ were chosen based on Fig.~3 of \cite{2021MNRAS.502.2970A}, in order to be broadly realistic for the MeerKAT system.

\subsubsection{Effect on foreground removal}
Fig.~\ref{figure:map_bs} shows the output intensity maps from the beam squint simulations generated using the MeerKAT survey specifications, after the application of the blind foreground removal procedure where we set the number of sources $n_s=5$ for the extended spin-0, 1, and 3 map-making, and $n_s=6$ for the spin-0 map-making. The difference in $n_s$ allows for a fair comparison, since the extended map-making should have reduced the number of contaminants by removing the systematic signal. Also plotted in Fig.~\ref{figure:map_HI_1} is the input HI cosmological signal for comparison. All of the maps have had the same mask applied to allow for simple comparison. The maps plotted here are for the $\sim$915.5MHz frequency band, but the results are generally similar for all frequency bands.

In Fig.~\ref{figure:map_random} we see the output of the simulation where a beam squint systematic that is non-smooth in frequency is present. The right panel shows that in the spin-0 map-making case the retrieved signal is clearly dominated by the beam squint leakage, and particularly by the $\eth I$ signal. Despite the fact that the $\eth I$ is smooth in frequency, the non-smooth $\zeta$ offset prevents the foreground removal from being able to remove this signal. The left panel shows that when using the extended spin-0, 1 and 3 map-making procedure, the retrieved signal is no longer subject to contamination by the systematic; the map-making has removed the leakage prior to the foreground extraction, allowing the HI cosmological signal to be retrieved successfully.

In contrast Fig.~\ref{figure:map_linear} shows that if the beam squint systematic is smooth in frequency, standard foreground removal can deal with the contamination, at least for the representative values of beam squint taken from Fig.~3 of \cite{2021MNRAS.502.2970A}. The contamination is dominated by the $\eth I$ leakage, however as shown in Fig.~\ref{figure:dIvsfrhosmooth} this component of the systematic is smooth, and so standard blind component separation methods can successfully remove it. As can be seen in Fig.~\ref{figure:dIvsfrhosmooth}, the $\eth P$ leakage is non-smooth in frequency, and as such can evade the standard foreground cleaning techniques. As expected, its contribution is very small compared to that of $\eth I$, and so the remaining contamination after the standard foreground removal procedure is insignificant. (Note that while this statement is true for the representative beam squint $\zeta$ systematic we have input, it is dependent on its size.) Consequently both the standard spin-0 and the extended spin-0, 1 and 3 map-making cases show successful recovery of the HI cosmological signal.

In Fig.~\ref{figure:map_bs} there is clearly a heavy loss of sky area, which stems from the cut made at the map-making stage based on how well-conditioned the map-making matrix is. As we are attempting to solve for several spin signals simultaneously, the conditioning becomes more sensitive to the crossing angle coverage. The large loss in sky area here is in part due to the scanning strategy being for a relatively small pathfinder survey, resulting in sparser crossing angle coverage, and also due to our choice to only consider a single dish. A natural benefit from a larger scale survey with additional dishes considered will be increased coverage, provided the scan strategy is designed to maximise crossing angle coverage given other constraints. We discuss this more in Appendix~\ref{section:Scanning Strategies}.

A given experiment will need to quantify its expected beam squint distribution in frequency. In most cases this will be smooth, and so the loss in survey coverage from performing the extended spin-0, 1 and 3 map-making likely outweighs the benefits, given that foreground removal should deal with the worst of the contamination in this case. Instead, the standard spin-0 map-making, which would retain the full survey area, would suffice without much detriment from the beam squint in this scenario. In some less common cases, a non-smooth in frequency beam squint can manifest, and in this case the leakage would be large enough that its removal via extended map-making would be a benefit that could outweigh the loss in survey area.

\begin{figure}
  \centering
    \subfloat[Random Beam Squint Simulation.\label{figure:c_l_random}]{\includegraphics[width=0.95\columnwidth]{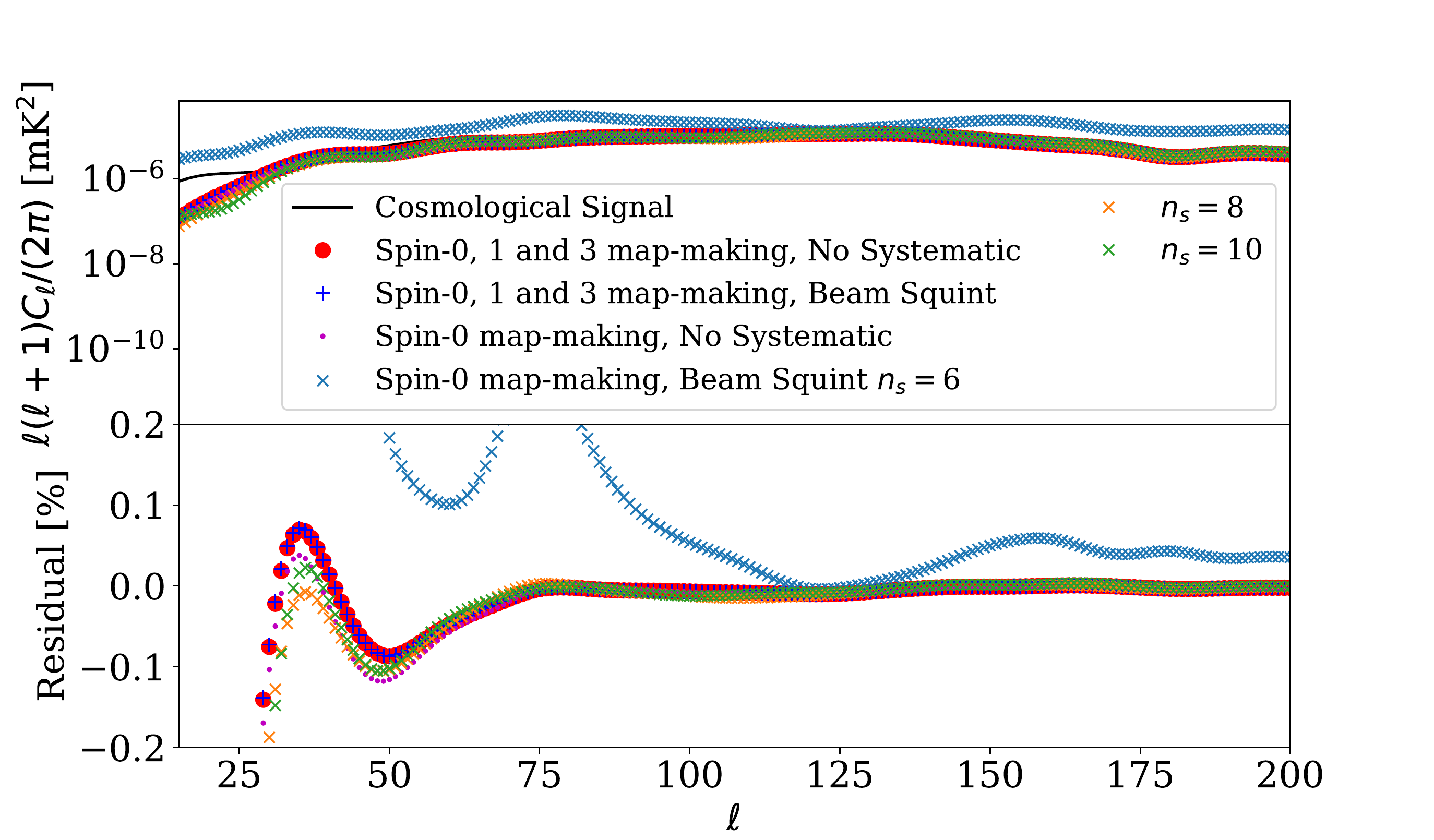}}
    \\
    \subfloat[Linear Beam Squint Simulation.\label{figure:c_l_linear}]{\includegraphics[width=0.95\columnwidth]{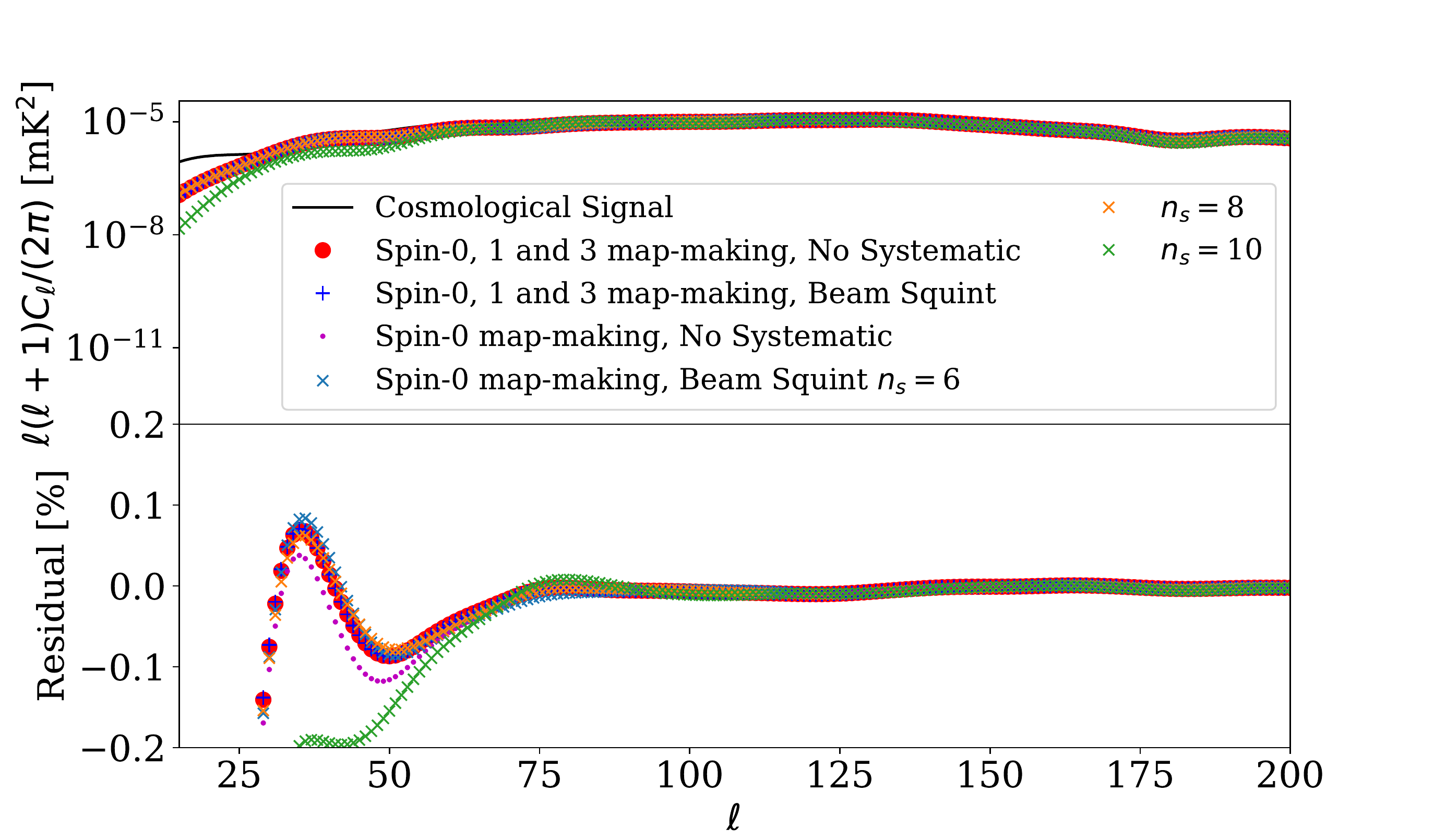}}
\caption{Foreground-cleaned angular power spectra at $\sim915.5$MHz. In both panels we see that the spectra output in the absence of systematics (red circles and pink dots) matches the input HI cosmological signal well for both cases of map-making. The slight disagreement at low-$\ell$ is a consequence of some loss of the cosmological signal during the foreground removal process. From Fig.~\ref{figure:c_l_random} we see that for a beam squint systematic that is non-smooth in frequency, implementing the standard spin-0 map-making used in intensity mapping and cleaning with $n_s=6$ (cyan crosses) results in contamination of the recovered HI angular power spectrum. For $n_s=8$ and 10, shown as orange and green crosses respectively, we see that the cleaning performs better. However, when implementing an extended map-making process solving for spin-0, 1, and 3 signals, we see that the contamination is removed (blue +s) in this case only using $n_s=5$ in the cleaning. Fig.~\ref{figure:c_l_linear} illustrates that in the case of systematics that are smooth in frequency, the standard foreground removal procedure will deal well with the contamination, although more aggressive cleaning (larger $n_s$) comes at the price of losing more cosmological signal at low-$\ell$ -- as is clear in the percentage residuals, taken between the cleaned data and the input cosmological signal (green crosses).}
\label{figure:c_ell_bs}
\end{figure}

\begin{figure}
  \centering
    \subfloat[Random Beam Squint Simulation.\label{figure:p_freq_random}]{\includegraphics[width=0.95\columnwidth]{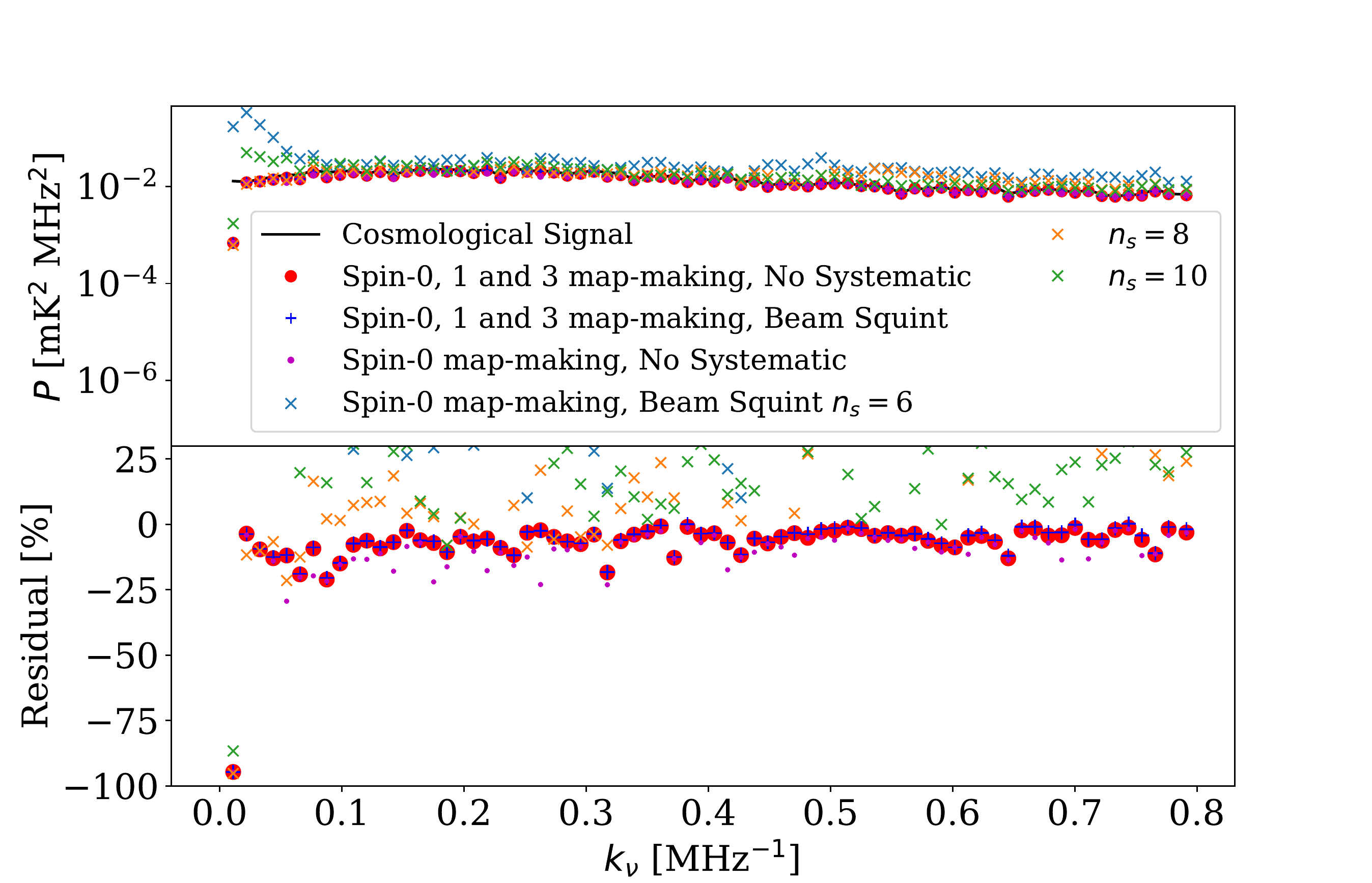}}
    \\
    \subfloat[Linear Beam Squint Simulation.\label{figure:p_freq_linear}]{\includegraphics[width=0.95\columnwidth]{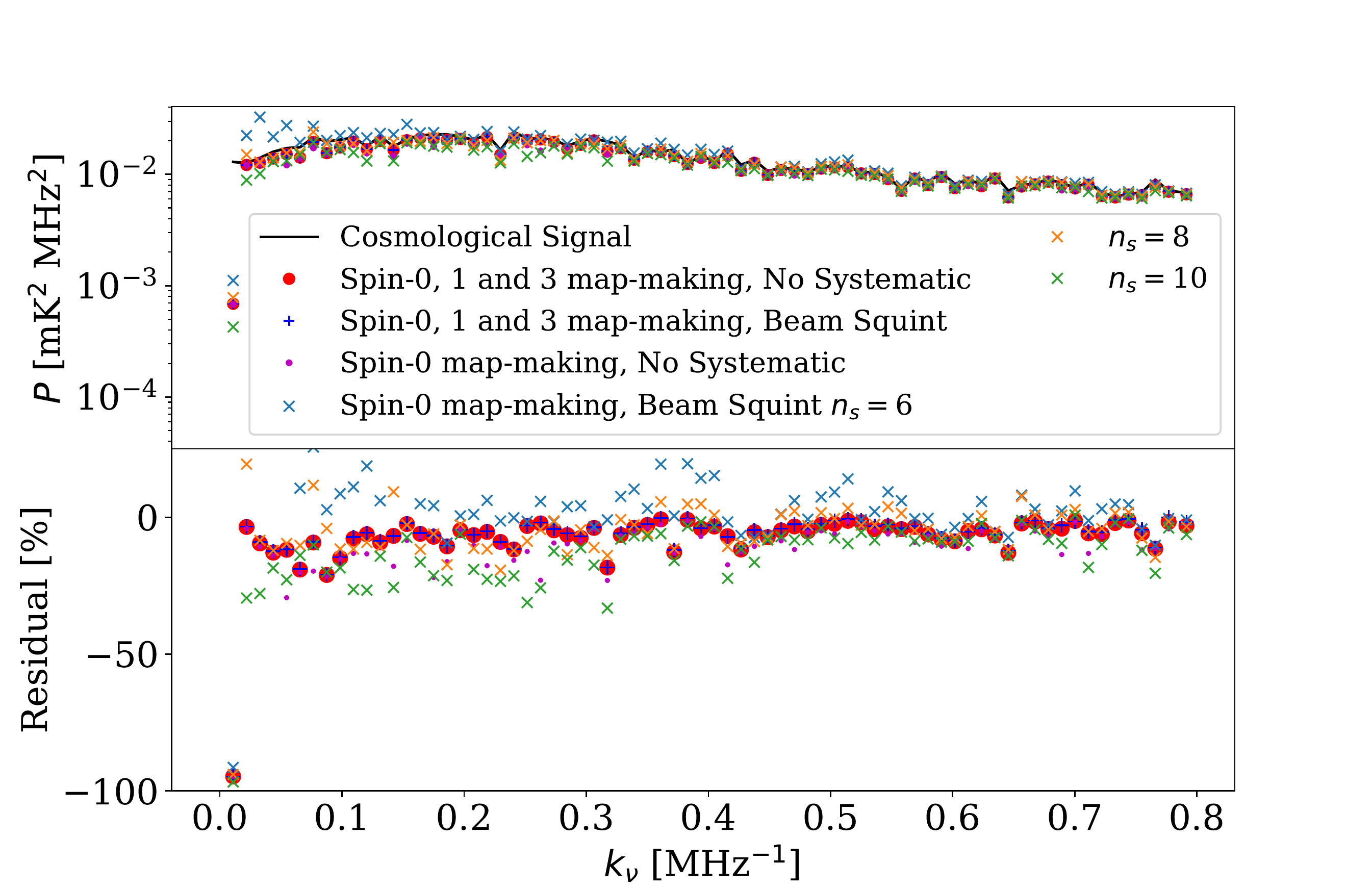}}
\caption{Foreground-cleaned radial power spectra. In both panels we see that the spectra output in the absence of systematics (red circles and pink dots) matches the input HI cosmological signal well for both cases of map-making. The percentage residuals in the lower panels of both Fig.~\ref{figure:p_freq_random} and \ref{figure:p_freq_linear}, taken between the cleaned data and the input cosmological signal, highlight the discrepancy at low-$k_{\nu}$ ($\lesssim 0.4$ MHz$^{-1}$), which is a consequence of some loss of the cosmological signal during the foreground removal procedure. From Fig.~\ref{figure:p_freq_random} we see that for a beam squint systematic that is non-smooth in frequency, implementing the standard spin-0 map-making results in a fairly large contamination to the retrieved HI signal for $n_s=6$ (cyan crosses). Increasing $n_s$ to 8 and 10, shown as orange and green crosses respectively, does reduce the contamination, but not completely. However when implementing the extended map-making process (blue +s) we see that the contamination is removed and the cleaning only requires a relatively small $n_s=5$. Fig.~\ref{figure:p_freq_linear} shows that a beam squint systematic that varies smoothly in frequency should be dealt with well by the foreground removal process itself when using a sufficiently large $n_s$ for the cleaning.}
\label{figure:p_freq_bs}
\end{figure}

Fig.~\ref{figure:c_ell_bs} and Fig.~\ref{figure:p_freq_bs} show, post foreground removal, the angular power spectra at $\sim915.5$ MHz and the radial power spectra respectively. Similarly to Fig.~\ref{figure:c_ell_g} and Fig.~\ref{figure:p_freq_g} of Section~\ref{section:Polarization Removal}, we see a slight disagreement at low-$\ell$ and low-$k_{\nu}$, stemming from some loss of the HI cosmological signal during the foreground removal process \citep{2020MNRAS.499..304C}.

Fig.~\ref{figure:c_l_random} and Fig.~\ref{figure:p_freq_random} show the case for a beam squint systematic that is non-smooth in frequency. The cyan, orange, and green crosses show the case where cleaning was performed with $n_s=6$, 8, and 10 respectively, where a beam squint systematic was included and we employed the standard spin-0 map-making usually used in intensity mapping surveys. When using $n_s=6$, the observed intensity signal is dominated by the systematic contamination, and the beam squint systematic-leaked signal has not been removed by the foreground cleaning due to its non-smooth dependence on frequency. For the larger $n_s$ cases we see that the cleaning does improve even for the standard spin-0 map-making with most of the contamination removed. However when implementing the extended map-making process, solving for spin-0, 1 and 3 signals prior to foreground removal, the output spectra (shown as blue +s) matches the input cosmological signal (black solid line) well even though only a relatively small number of sources were included in the cleaning process ($n_s=5$). The extended map-making removes the systematic contamination, allowing the foreground removal to successfully recover the HI cosmological signal with less aggressive cleaning.

In contrast Fig.~\ref{figure:c_l_linear} and Fig.~\ref{figure:p_freq_linear} illustrate that, in the case where we have a beam squint systematic that is smooth in frequency, the standard foreground removal will deal with most of the contamination even when using the standard spin-0 map-making and cleaning with a relatively low $n_s=6$ (shown as cyan crosses). For both the spin-0 and the spin-0, 1, and 3 map-making cases, the output spectra for the simulations including the systematic line up well with the output spectra from the simulations with no systematics. However as evidenced by the percentage residual in the lower panel of Fig.~\ref{figure:c_l_linear}, using a more aggressive cleaning with $n_s=10$ (green crosses) comes at the price of losing more cosmological signal at low-$\ell$. Additionally, in Fig.~\ref{figure:p_freq_linear} we see that for $n_s=6$, the spin-0 map-making case deviates slightly more from the cosmological signal. This is due to the non-smooth in frequency $\eth P$ leakage, however for larger $n_s$ it is close to insignificant for the representative levels of beam squint used for the smooth linear variation in frequency case.

\subsubsection{Summary}
In summary, we know that there is a dependence of beam squint on frequency and it is important to understand how smooth that dependence is to know whether the contamination will be removed by standard blind foreground removal, or whether extended map-making is required. We expect that in most cases the squint will vary smoothly with frequency, and have demonstrated that the standard foreground removal will deal with the contamination without requiring the extended map-making in this case. This is despite the systematic having a non-smooth component from the $\eth P$ leakage, due to its contribution being sufficiently small that there is very little contamination remaining after the foreground cleaning.

We have also demonstrated that in the less common scenario where the systematic is non-smooth in frequency, the standard approach of spin-0 map-making will result in the systematic heavily contaminating the observed intensity signal when cleaning with $n_s=6$, but when using larger $n_s$ the cleaning does a better job of removing the contamination. We have shown that this systematic leakage can be removed via map-making prior to the application of the foreground removal process. When additional spin-1 and 3 signals are solved for at map-making, the contamination is removed from the observed intensity signal, which allows the cleaning process to work for a lower number of sources -- e.g. $n_s=5$, as was used in this case. As such, if the beam squint systematic is non-smooth in frequency for an experiment, then the extended map-making method could prove an important technique for removing the resulting contamination, making the blind foreground removal process easier.

\subsection{General application and limitations}
\label{section:General application and limitations}
We have shown that the inclusion of additional spins in simple binned map-making is a viable technique to remove systematics from intensity mapping surveys. This offers a simple extension to the map-making techniques already employed by autocorrelation intensity mapping experiments, but requires careful thought about scanning strategies to ensure sufficient redundancy in crossing angle coverage. Importantly, this technique allows us to exploit the spin structure of troublesome systematics that leak signals such as the polarized foregrounds, which are non-smooth in frequency and so are not removed effectively by blind component separation methods.

The choice of which additional spin(s) to include in the map-making process will depend on the systematics that one is attempting to remove. Table~\ref{tab:systematics} gives some examples of systematics that could potentially be disentangled from the cosmological signal by this technique, provided the map-making is performed on the data over timescales on which the systematic is stable. This list is not exhaustive; any systematics which are stable in time and have a well defined spin dependence could be included in this list. The choice of additional spins to include in the map-making will be dictated by which systematics are of most concern to a given experiment. As such, the technique can be tailored to be most useful to a specific survey.

Another important feature is that this process does not depend on the source of the systematic, but rather only on the spin dependence of the leaked signal. For instance, regardless of the cause, if the spin-2 on-sky polarization signal is leaked into the observed intensity, then it may be disentangled in this way. Provided one has sufficient redundancies in the scanning strategy, and that the systematic is stable in time, then making use of Eq.~\ref{eq:Spin 2 3x3 Map-making} to solve for both a spin-0 and spin-2 signal in the map-making process will isolate the intensity signal from the polarization leakage.

There are, of course, a number of limitations to the application of this technique which we shall discuss here. The method's reliance on the leaked signals being stable in time is the primary limitation, with many systematics of concern to upcoming surveys potentially having a strong time dependence. It would be difficult to treat such signals within the framework presented here. Other techniques, such as periodically observing a calibration source during a scan, timestream filtering, and destriping, would be required to remove such systematics. Note, however, that if the systematics drift slowly over the duration of a full survey, then map-making to remove them could potentially be performed on TOD split into shorter timescales over which the systematic is stable. The resulting observed intensity maps could then be stacked before applying foreground removal. This approach would rely on there being sufficient crossing angle coverage in the regions hit by the scan during the shorter timescale chunks to facilitate the extended map-making.

\begin{table}
	\centering
	\caption{A set of systematics that can be modelled and mitigated by the methods presented in this paper. The column labelled ``$\tilde{h}_{n}$ Coupling'' gives the orientation quantity of interest that could be suppressed with a well designed scanning strategy. The column labelled ``Spin'' corresponds to the additional spin that could be solved for during the map-making process to disentangle the systematic from the true signal. This list is by no means exhaustive, but gives an indication of the systematics these techniques are appropriate for.}
	\label{tab:systematics}
	\begin{tabular}{lcc} % four columns, alignment for each
		\hline
		Systematic & $\tilde{h}_{n}$ Coupling & Spin\\
		\hline
		Beam Squint (Intensity derivative leakage) & $\tilde{h}_{\pm1}$ & 1\\
		Beam Squint (Polarization derivative leakage) & $\tilde{h}_{\pm 1}$ and $\tilde{h}_{\pm 3}$ & 1 and 3\\
		Beam Ripple (Polarization leakage) & $\tilde{h}_{\pm2}$ & 2\\
		Gain Mismatch (Polarization leakage) & $\tilde{h}_{\pm2}$ & 2\\
		Imperfect Alignment (Polarization leakage) & $\tilde{h}_{\pm2}$ & 2\\
		\hline
	\end{tabular}
\end{table}

This highlights another limitation of the technique. By extending the map-making to solve for additional signals other than the spin-0 intensity, the map-making process becomes more complicated (see e.g. Eqs.~\ref{eq:Spin 2 3x3 Map-making}, \ref{eq:5x5mapmaking}, and \ref{eq:Spin 1 3x3 Map-making}). The presence of the additional spins makes the process far more dependent on having a well-designed scanning strategy, since the matrix inversion requires sufficient redundancies to be well-conditioned. For a realistic scan strategy, a data cut would be required on ill-conditioned pixels, which would result in the loss of some of the observed field. As such, in order to use the extended map-making technique, careful thought will be required for design of the scanning strategy for upcoming surveys. We comment on this more in Appendix~\ref{section:Scanning Strategies}.

This also relates to the final limitation we shall discuss. One may be tempted to arbitrarily extend the map-making to solve for many spin signals simultaneously, in the hope of covering all possible relevant leaked signals. Unfortunately this is not possible, as the more spins that are included when map-making, the less well-conditioned the inversion of the map-making matrix is likely to be. The introduction of additional spins means that good crossing angle coverage in the scanning strategy becomes even more important. See Appendix~\ref{section:Scanning Strategies} for further information.

\section{Conclusions}
\label{section:Conclusion}
We have adapted the formalism from \cite{Wallisetal2016}, \cite{2021MNRAS.501..802M}, and \cite{2021arXiv210202284T} and applied it in the context of autocorrelation (`single-dish') intensity mapping surveys. This formalism classifies instrumental systematics according to their spin properties, and therefore acts as a framework for studying and mitigating these systematics, including those that cause polarization leakage.
The key results are as follows:
\begin{enumerate}
    \item We have provided a simple approach to modelling systematics with well-defined spin dependence. The key equation, Eq.~\ref{eq:SpinSyst}, gives a general description of an observed spin-0 field and how signals of other spin can leak into it. Combining this with the simple intensity mapping setup of Section~\ref{section:Formalism} allows for the leakage into the observed intensity signal due to a number of systematics to be described. This can be used in conjunction with prior knowledge of the scanning strategy and instrumental parameters to estimate the expected level of certain systematics, and how well they are mitigated through the crossing angle coverage achieved by the scan.
    
    \item We showed that a straightforward extension of the simple binned map-making method -- solving for additional spin fields -- provides a method that disentangles such systematics from the cosmological signal. The spins to include in map-making should be tailored to the systematic of most concern to a given instrument. We have shown that this map-making offers an approach to remove sources which are troublesome to existing foreground removal techniques, making the subsequent cleaning process easier, at least to the level of realism incorporated in our simulations so far. Specifically, the removal of the systematic signals via map-making prior to the application of foreground cleaning reduces the number of sources required by the component separation methods to successfully recover the cosmological signal. The limitations of this approach were discussed in Section~\ref{section:General application and limitations}; we again highlight the key issue in its application, which is the reliance on the systematic signals being stable in time.
    
    \item We have shown that polarization leakage contaminating the intensity signal can be removed by map-making for both spin-0 and spin-2 signals simultaneously, using an effective gain mismatch systematic as an example. For the case presented when implementing standard spin-0 map-making, the foreground removal procedure struggled to deal with the contamination from polarization even when using a large number of sources ($n_s=10$). However, when implementing the extended spin-0 and 2 map-making to remove the polarization leakage, prior to applying blind component separation, the subsequent foreground removal was successful for a comparatively modest $n_s=5$.
    
    \item We have also shown a case study where contamination of the intensity signal can be sourced by systematics of different spin, even in the case where there is no polarization leakage. This technique removes the systematic even when the contaminating signals are non-smooth in frequency -- signals that are not readily removed by blind component separation methods.
    
    \item Though we did not expand on it much in this proof of concept, we proposed an alternative use for the extended map-making method: a ``blind'' search could be performed to discover the spin of the systematics that are most contaminating the data. Given that we reasonably expect only a discrete number of spin signals to manifest in the data, we anticipate that a ``blind'' diagnosis of systematics could be implemented, by making maps that include progressively more spin fields in addition to the expected spin zero intensity. The resulting reconstructed fields can then be inspected for any signs of systematic signatures that could be corrupting the cosmological signal. We point out that this proposed method will suffer from similar limitations, such as requiring that the systematics be stable in time over the data incorporated. We present a first look at this possibility in Appendix~\ref{section:Blind Diagnosis}, but leave to future work a detailed examination of this approach.
\end{enumerate}
We have shown a proof of concept here for how a spin-based map-making technique can be used to deal with systematics in intensity mapping surveys that would otherwise be difficult to filter out. The general processes outlined here, particularly the map-making, should be beneficial to a number of planned intensity mapping surveys, e.g. with MeerKAT and SKAO-MID, and their relative simplicity should facilitate their implementation for future experiments. Crucially, we expect the tools shown here to work for any kind of polarization leakage that is caused by instrumental non-idealities, providing they can be written in the form of Eq.~\ref{eq:decomp}. These tools now need to be applied to increasingly realistic and complex simulations, in order to understand the full ramifications of using them on real data.

\section*{Acknowledgements}
We are grateful to Jingying~Wang for useful comments and help with the MeerKAT survey specifications. The authors acknowledge Isabella Carucci, Melis Irfan, Mario Santos, Yi-Chao Li, Stuart Harper, Laura Wolz, Richard Battye, Keith Grange, and the MeerKAT collaboration for helpful discussions. NM is supported by a STFC studentship. DBT acknowledges support from Science and Technology Facilities Council (STFC) grants ST/P000649/1, ST/T000414/1 and ST/T000341/1. This result is part of a project that has received funding from the European Research Council (ERC) under the European Union's Horizon 2020 research and innovation programme (Grant agreement No. 948764; PB). PB also acknowledges support from STFC Grant ST/T000341/1.

%%%%%%%%%%%%%%%%%%%%%%%%%%%%%%%%%%%%%%%%%%%%%%%%%%
\section*{Data Availability}
The data underlying this article will be shared on reasonable request to the corresponding author. The codes used will be made available at \url{https://github.com/NiumCosmo/Spin-Based_Map-Making_IM}.



\bibliographystyle{mnras}
\bibliography{IM} % if your bibtex file is called example.bib

%\begin{thebibliography}{99}
%\end{thebibliography}

%%%%%%%%%%%%%%%%%%%%%%%%%%%%%%%%%%%%%%%%%%%%%%%%%%

%%%%%%%%%%%%%%%%% APPENDICES %%%%%%%%%%%%%%%%%%%%%

\appendix

\section{Scanning Strategy}
\label{section:Scanning Strategies}
We adopt the scanning strategy put forth in \cite{2020arXiv201113789W} in this paper, however this was in fact a path finder survey and a full survey design would likely have both increased integration time as well as exploit further redundancies in the scanning strategy. We have demonstrated in this work that the extended map-making procedure can remove systematic leakage, but with an associated loss of coverage area due to cutting pixels with ill-conditioned map-making matrices. This is particularly evident in the case of the beam squint systematic where we use Eq.~\ref{eq:5x5mapmaking} to solve for spin-0, 1, and 3 signals simultaneously.

This is mainly due to the poor crossing angle coverage (particularly in the edge pixels) born out of this being a relatively small path finder survey. This should be naturally remedied with the more complex scanning strategy which would likely be in place for a full survey e.g. MeerKLASS \citep{2017arXiv170906099S}.

The $|\tilde{h}_n|^2$ act as a metric of the crossing angle coverage and give an indication both of how well the scan strategy should suppress systematics of spin-n -- they are also tied to how well including a spin-n signal in map-making should perform. These quantities can even act as and indicator of what elevation or field would be useful to include in a scanning strategy. If one calculates the integral of these quantities over the planned observed patch and minimises them by varying either the field location or elevation of the telescope this will give an indication of the best setup to suppress systematics.

The introduction of additional spins to map-making makes the process far more dependent on having a well designed scanning strategy. For a given scan the more spins that are introduced into map-making the less well conditioned the inversion of the map-making matrix is likely to be. However provided the scan has sufficient crossing angle coverage the map-making should behave well enough even when solving for multiple spins. There will also be a dependence on the chosen pixel size, coarser pixels will naturally incorporate more hits for a given scanning strategy. This will allow more data to be included in the map-making step allowing for improved conditioning although with a sacrifice to the angular resolution of the output maps.

\begin{figure*}
  \centering
    \subfloat[Spin 0 and 2 map-making coverage. Left Panel: MeerKAT scan strategy. Right Panel: MeerKAT + Extra Low Elevation.\label{figure:spin0_2cov}]{\includegraphics[width=1.7\columnwidth]{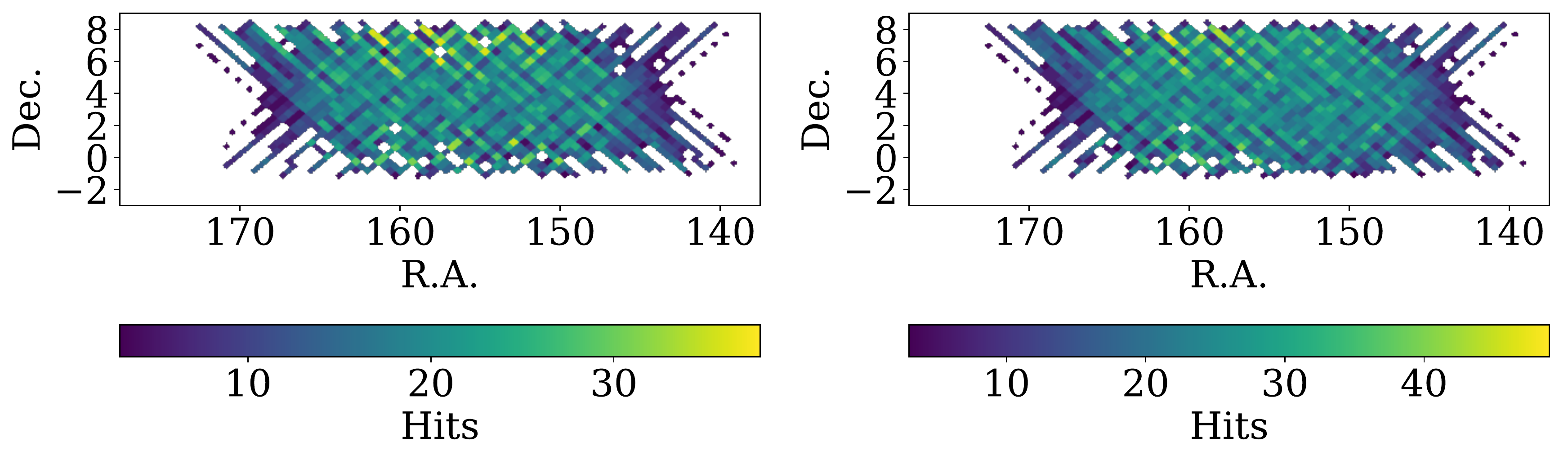}}
    \\
    \subfloat[Spin 0, 1 and 3 map-making coverage. Left Panel: MeerKAT scan strategy. Right Panel: MeerKAT + Extra High Elevation.\label{figure:spin0_1_3cov}]{\includegraphics[width=1.7\columnwidth]{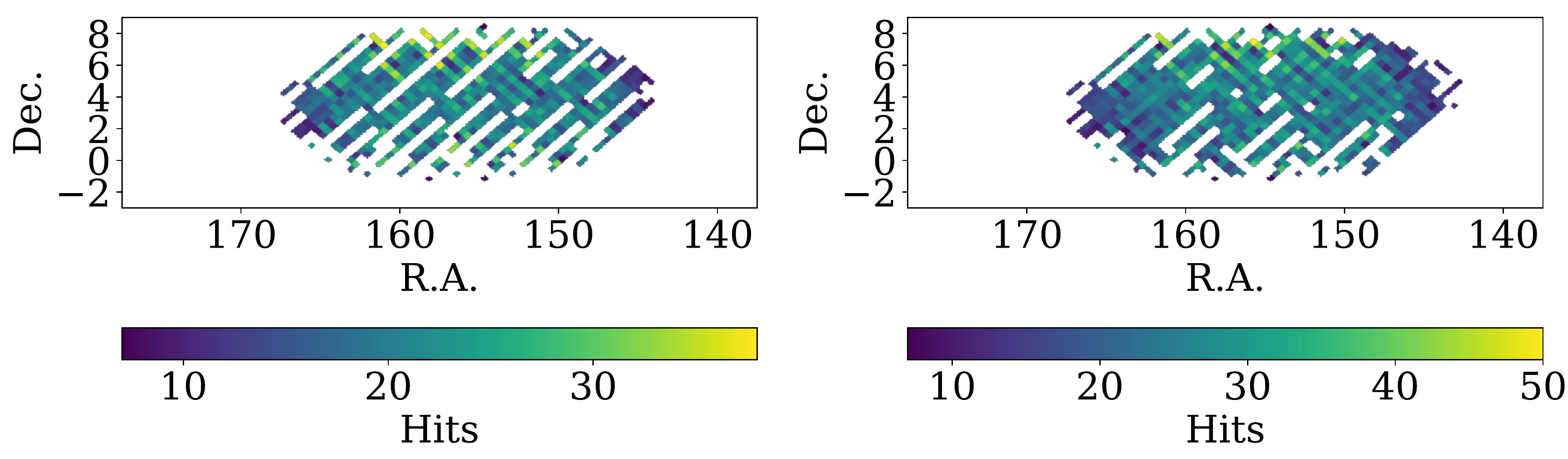}}
    \\
    \subfloat[Spin 0, 1 and 3 map-making coverage. Left Panel: MeerKAT scan strategy. Right Panel: MeerKAT + Extra Low Elevation.\label{figure:spin0_1_3cov_low}]{\includegraphics[width=1.7\columnwidth]{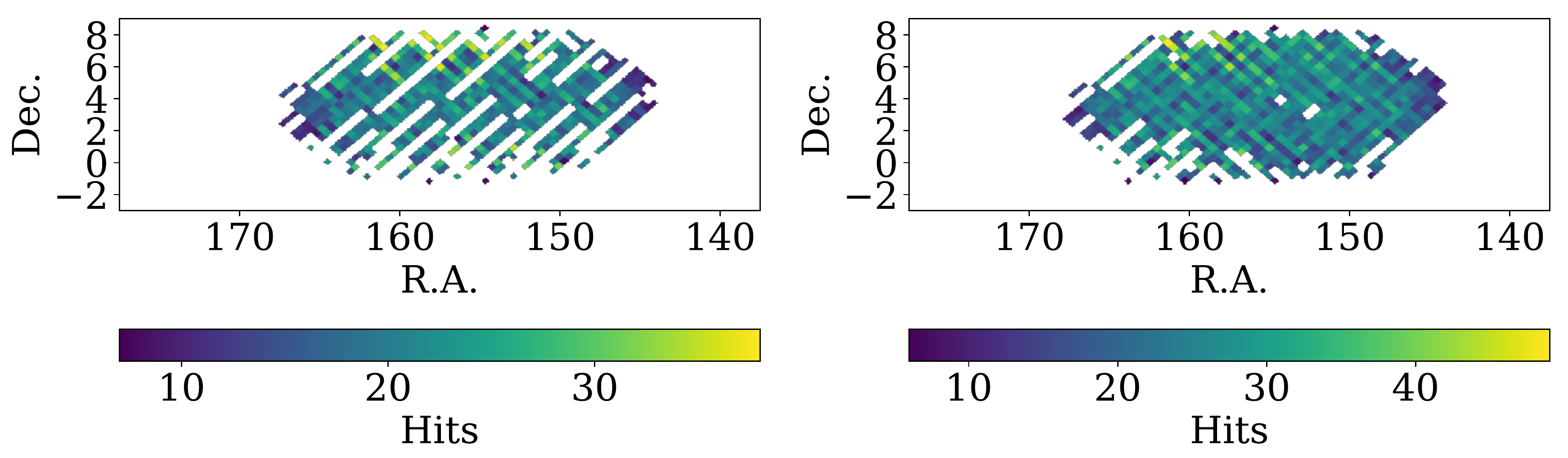}}
\caption{Panel \ref{figure:spin0_2cov} shows the resulting hitmaps following the cut made based on the conditioning of the map-making for spin-0 and 2 signals. Comparing the coverage attained using the MeerKAT survey of \protect\cite{2020arXiv201113789W} and the case where two further observations of the field are made at rising and setting with the telescope elevation set to the optimal elevation to minimise $|\tilde{h}_2|^2$ ($39.0^{\circ}$ for this field) we see a slight increase in the number of pixels that survive the cut. Panel \ref{figure:spin0_1_3cov} shows the resulting hitmaps following the cut made based on the conditioning of the map-making for spin-0, 1 and 3 signals. Comparing the coverage attained using the MeerKAT survey and the case where two further observations of the field are made at rising and setting with the telescope elevation set to $49.0^{\circ}$ we see a large increase in the number of pixels that survive the cut. Panel \ref{figure:spin0_1_3cov_low} shows the resulting hitmaps following the cut made based on the conditioning of the map-making for spin-0, 1 and 3 signals. Comparing the coverage attained using the MeerKAT survey and the case where two further observations of the field are made at rising and setting with the telescope elevation set to $39.0^{\circ}$ we see the number of pixels that survive the cut increases even more.}
\label{figure:coverage}
\end{figure*}

Fig.~\ref{figure:spin0_2cov} shows that the MeerKAT scan strategy is already close to optimal for the spin-0 and 2 map-making with most pixels in the field surviving the map-making cut. The optimal elevation to minimise the $|\tilde{h}_2|^2$ quantity for this specific field is $39.0^{\circ}$ (note that the optimal elevation will vary for different choices of field), and when implementing an extra two scans at rising and setting using this elevation we see a modest improvement in the number of pixels that survive the cut.

Fig.~\ref{figure:spin0_1_3cov} shows that the MeerKAT scan strategy is sub-optimal for spin-0, 1 and 3 map-making with many pixels being removed during the map-making cut. We also show the case where an additional two scans are performed at rising and setting at an elevation of $49.0^{\circ}$ -- this was calculated to be optimal for minimising the $|h_3|^2$ quantity for the field. In this case we see a large improvement in the number of pixels that survive the cut. The edges of the field are still cut in this case due to a sparse crossing angle coverage in those regions.

The crossing angle coverage achieved by the scanning strategy at the field edges is worse than at the centre and as such the map-making is less well conditioned there. Incorporating further elevations could help improve the crossing angle coverage as they would provide a scanning strategy with greater redundancies.

In contrast to $|h_3|^2$ the $|h_1|^2$ quantity for the field is minimised for extremely low elevations. This is the other quantity which is important to suppress for spin-0, 1 and 3 map-making, given the spin-1 dependence. In Fig.~\ref{figure:spin0_1_3cov_low} we show the effects of including a relatively low elevation of $39.0^{\circ}$ on the spin-0, 1 and 3 map-making -- though this elevation is still not optimal for the suppression of $|h_1|^2$ we see that its inclusion to the scanning strategy still vastly improves the number of pixels that survive the map-making cut. As such when including multiple additional spins at map-making it is important to examine each relevant $|h_n|^2$ quantity together to see what scanning strategy design will best aid in map-making overall.

Fig.~\ref{figure:coverage} demonstrates that with slight modifications to the MeerKAT scanning strategy of \cite{2020arXiv201113789W} the crossing angle coverage required for the extended map-making to be well conditioned can be achieved in most of the field. This supports the fact that the coverage issues for the spin-0, 1 and 3 map-making are due to the pathfinder nature of the survey, and should be fixed by a larger survey design.

\section{Blind Diagnosis of Systematics} \label{section:Blind Diagnosis}
We only expect a discrete number of spin signals from systematic leakage to manifest in the data. As such it is conceivable that a blind diagnostic approach can be implemented to determine which spin signals are contaminating the observations.

\begin{figure*}
  \centering
  \includegraphics[width=1.9\columnwidth]{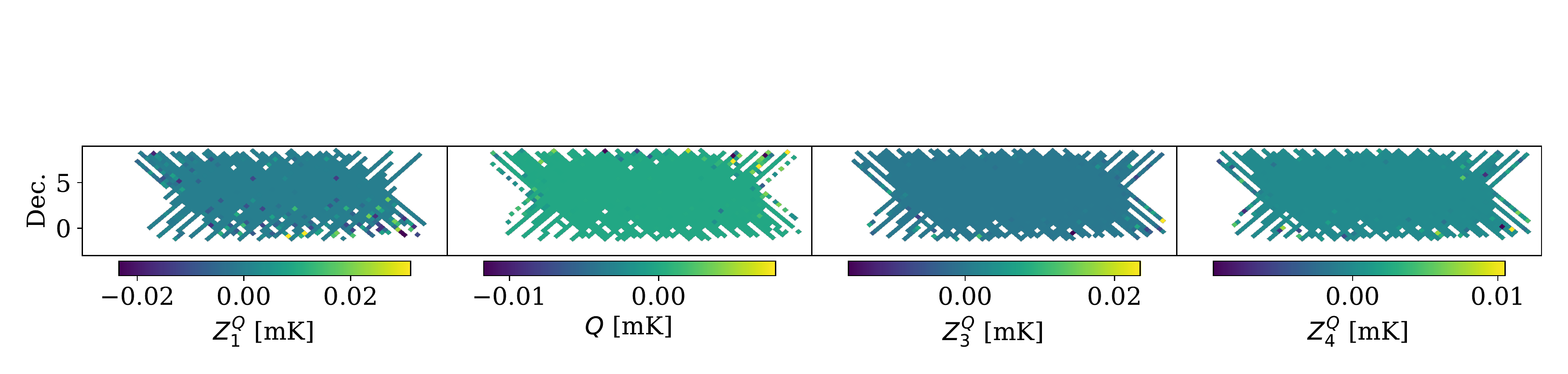}
  \includegraphics[width=1.9\columnwidth]{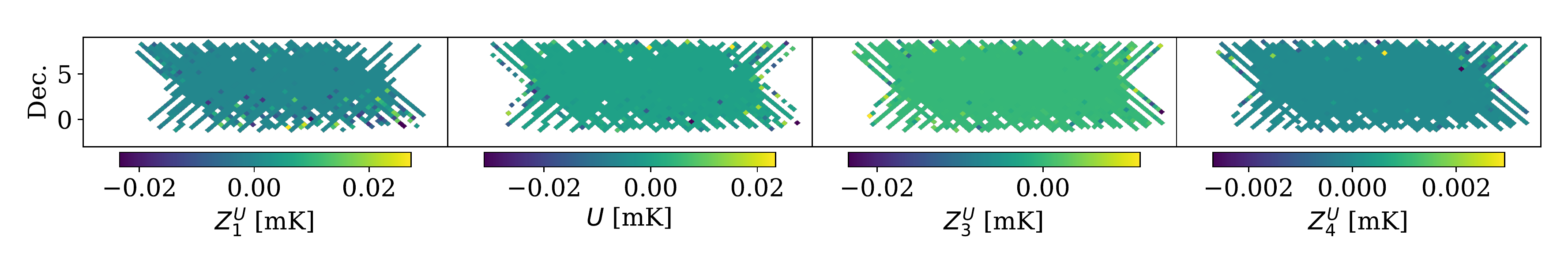}
  \includegraphics[width=1.9\columnwidth]{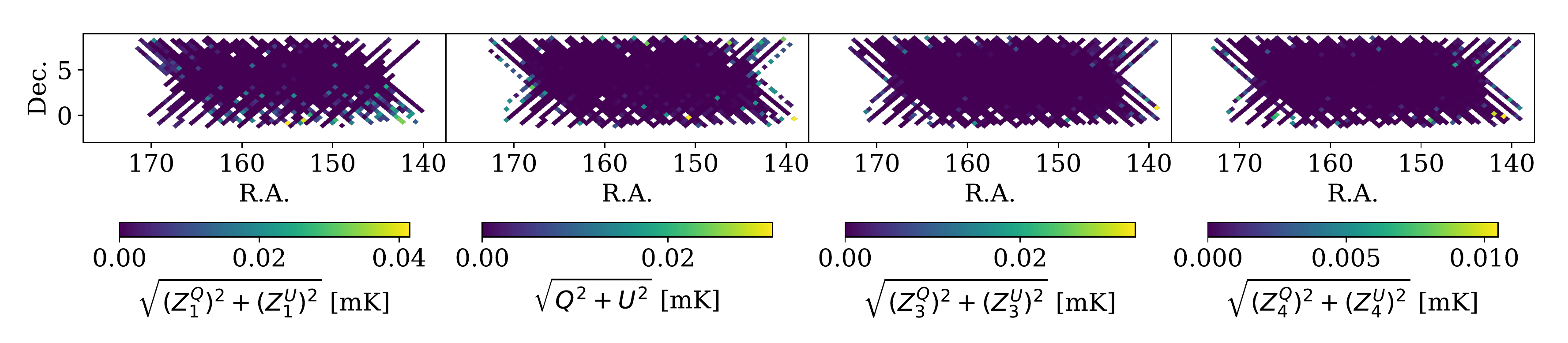}
\caption{Maps at a single frequency band of $~915.5$MHz prior to any foreground cleaning for a simulation with no systematic. The first column shows the reconstructed spin-1 signal when using spin-0 and 1 map-making, the second column shows the reconstructed spin-2 signal when using spin-0 and 2 map-making, the third column shows the reconstructed spin-3 signal when using spin-0 and 3 map-making, and the fourth column shows the reconstructed spin-4 signal when using spin-0 and 4 map-making. We see that all the data is very close to zero for the spin-1, 2, 3, and 4 fields as expected since the only signal present should be the spin-0 intensity if no systematics are affecting the observations. There are some edge effects at the extremities of the survey area resulting from the map-making not behaving quite as well there -- this is due to a reduced number of crossing angles. However the majority of the values in each field are of order $10^{-6}$ and less.\label{figure:blindnosyst}}
\end{figure*}

One method to do this is to successively make maps that include a single additional spin field (as well as spin-0) and examine the reconstructed spin fields for any signs of systematic signatures that could be corrupting the cosmological signal. We present a simple example of this using a toy simulation containing a gain mismatch of 1\%, as was previously studied in Sects.~\ref{section: Spin Characterisation of Systematics} and \ref{section:Map-Making}. The signals present are thus the expected spin-0 intensity signal, including both the cosmological signal and foregrounds, and a spin-2 signal from polarization leakage due to the systematic. For context we also present the results in a case with no systematics present.

In Fig.~\ref{figure:blindnosyst} we show a an example of attempting a blind diagnosis with maps made at a single frequency channel at $915.5$~MHz, for a simulation with no systematic present, prior to the blind foreground cleaning step. We show the output from implementing extended map-making including spin-0 and one additional signal of spin-1 (first column), 2 (second column), 3 (third column), and 4 (fourth column). With no contributions from systematics, the only signal present is the spin-0 intensity, and as expected we see that the reconstructions of the spin-1, 2, 3, and 4 fields from map-making are very close to zero, indicating no contamination. This also shows that, as a consequence of solving for the spin-0 signal as well as the additional spin during map-making, that no leakage is occurring from the spin-0 intensity signal into the other spin signals -- so the spin-0 signal itself will not influence the outcomes of the diagnostic approach.

\begin{figure*}
  \centering
  \includegraphics[width=2\columnwidth]{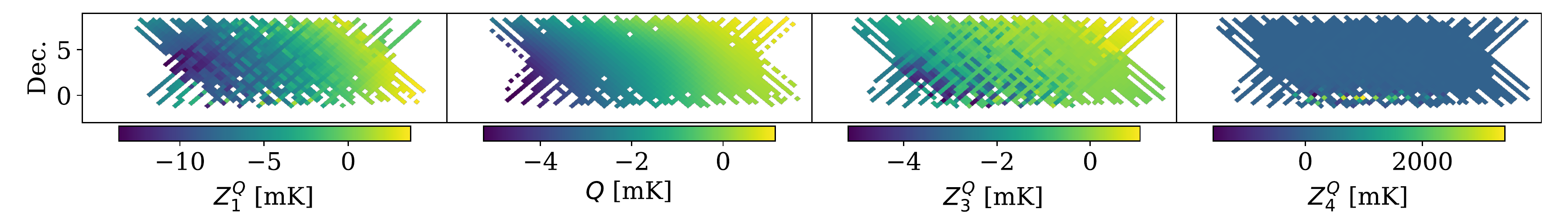}
  \includegraphics[width=2\columnwidth]{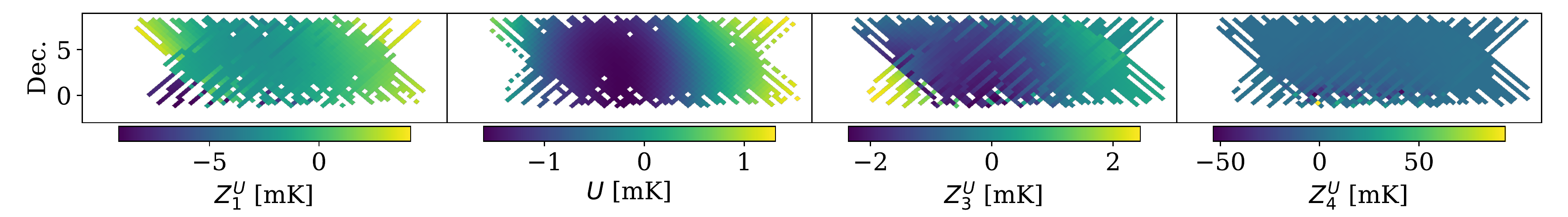}
  \includegraphics[width=2\columnwidth]{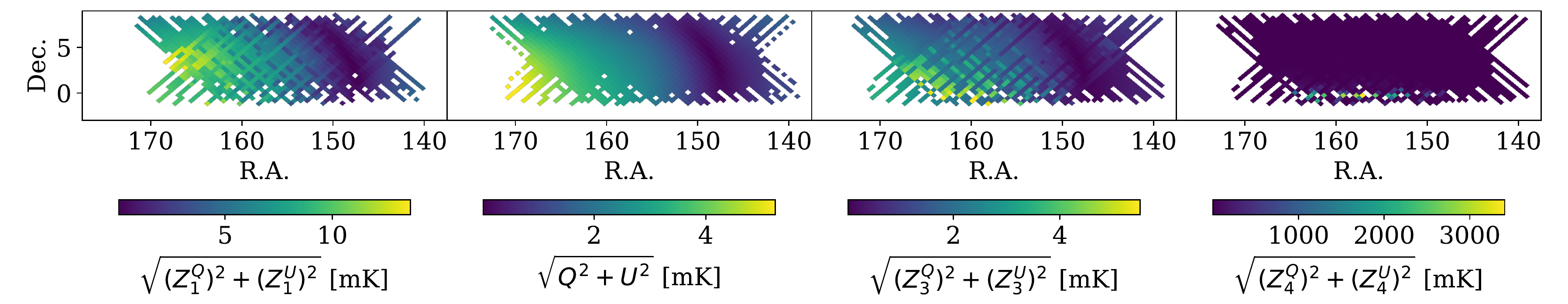}
\caption{Maps at a single frequency band of $~915.5$MHz prior to any foreground cleaning for a simulation including an effective gain mismatch. The first column shows the reconstructed spin-1 signal when using spin-0 and 1 map-making, the second column shows the reconstructed spin-2 signal when using spin-0 and 2 map-making, the third column shows the reconstructed spin-3 signal when using spin-0 and 3 map-making, and the fourth column shows the reconstructed spin-4 signal when using spin-0 and 4 map-making. The smooth structure of the fields in the second column suggests it is an actual spin-2 systematic signal rather than some other signal leaking into the spin-2. The spin-1, 3, and 4 fields all have significant structure present, likely from the scanning strategy, indicating they are likely spurious signals due to leakage from signals of other spin, and as such may be disregarded as the dominant systematic. The spin-4 case is particularly large, likely because the scan strategy is sub-optimal to suppress the leakage to that spin signal. \label{figure:blind}}
\end{figure*}

In Fig.~\ref{figure:blind} we show a simple example of attempting a blind diagnosis with maps made in the same frequency channel, but now in the presence of an effective gain mismatch. The columns are the same as in the previous figure. We see that the maps in the second column have smooth structure that is not consistent with zero. This is indicative of the presence of an actual spin-2 systematic signal, rather than some other signal leaking into spin-2 (i.e. there is no scan structure present that is suggestive of leakage). The other columns, showing the contributions to spin-1, 3, and 4, all have significant structure present related to the scanning strategy, which indicates contamination by signals of other spin being leaked by the relevant $\tilde{h}_n$ scan terms. The spin-4 case is particularly large, likely because the scanning strategy is far from optimal to suppress the leakage to that spin.

This reasoning would point to the likely dominant contamination being from a spin-2 systematic signal. We of course know this to be the case for this contrived example, however the process outlined should work for a truly blind case. The simplistic case we present here demonstrates the principle of diagnosing systematics; in future work we intend to make this procedure more rigorous by devising a diagnosis metric based on the $\tilde{h}_n$ scan quantities.

\begin{figure*}
  \centering
  \includegraphics[width=1.8\columnwidth]{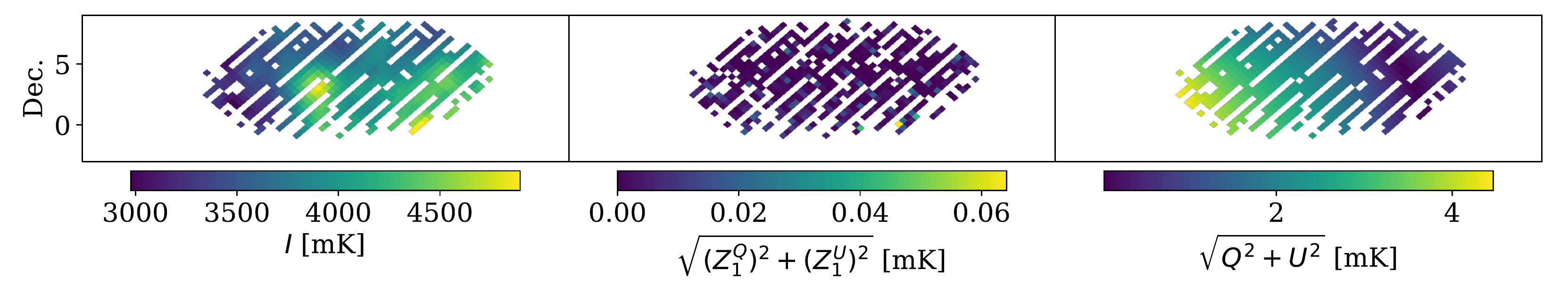}
  \includegraphics[width=1.8\columnwidth]{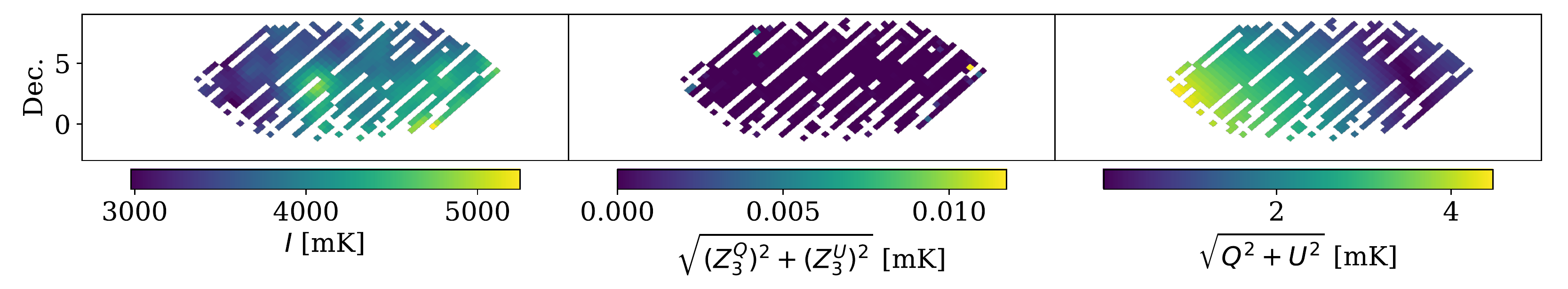}
  \includegraphics[width=1.8\columnwidth]{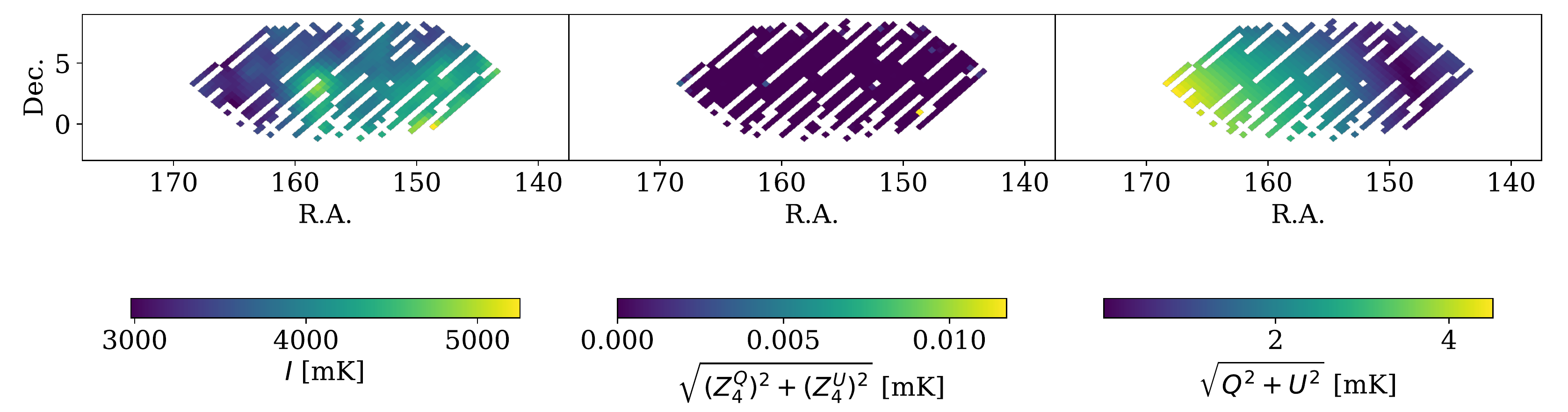}
\caption{Maps at a single frequency band of $~915.5$MHz prior to any foreground cleaning for a simulation including an effective gain mismatch. Row 1 shows the results of extended map-making including spin-0, 2, and 1 signals, row 2 for spin-0, 2, and 3 signals, and row 3 for spin-0, 2, and 4 signals. The first column shows the output intensity, the second column shows the output spin-n signal ($n \in \{1,3,4\}$), and the third column shows the output spin-2 $Q$ and $U$ signals. The spin-n signals are effectively zero since there is no systematic with that spin present, whereas the spin-0 intensity and spin-2 systematic show clear smooth signals thus confirming that there is a spin-2 systematic signal present.}
\label{figure:blind2}
\end{figure*}

Following this step, we can also check to see if this deduction is likely correct by performing further extended map-making, by always solving for a spin-0 and 2 signal along with one additional spin signal. Fig.~\ref{figure:blind2} shows this case with the map-making in row 1 solving for spin-0, 2, and 1 signals, row 2 solving for spin-0, 2, and 3 signals, and row 3 solving for spin-0, 2, and 4 signals. The first column shows the output intensity, the second column shows the output spin-$n$ signal ($n \in \{1,3,4\}$), and the third column shows the output spin-2 $Q$ and $U$ signals. For ease we display these together as $\sqrt{(Z_n^Q)^2+(Z_n^U)^2}$ and $\sqrt{Q^2+U^2}$ as this gives all the required information for this particular analysis. The additional spin-$n$ signals have effectively been reconstructed to zero, since there is no systematic with that spin present, while the spin-0 and spin-2 fields show obvious smooth signals. As per the reasoning above, this confirms that the systematic is of spin-2 origin. As a consequence of including three spins in the map-making, we see a reduction in the survey area however. This shows that one cannot arbitrarily extend the map-making to solve for many spin signals simultaneously in the hope of covering all possible relevant leaked signals -- the consequence of doing so would be less well behaved map-making with an associated cost of reduced survey area. As such one should tailor the map-making to include as few spins as possible by isolating which spin signal represents the dominant contaminant. In this simple case, the ideal map-making would be to solve for spin-0 and 2 only.

%\textcolor{red}{We have presented a simplistic case here that serves as a simple proof of concept of the diagnostic procedure. However this will require more detailed analysis in future work. We also note that the same limitation as earlier applies that the systematics must be stable in time over the data incorporated.}

%%%%%%%%%%%%%%%%%%%%%%%%%%%%%%%%%%%%%%%%%%%%%%%%%%

% Don't change these lines
\bsp	% typesetting comment
\label{lastpage}
\end{document}